\documentclass[preprint]{ptephy_om}

\preprintnumber{KYUSHU-HET-328} %%% %%% Insert preprint number here

\usepackage{graphics}

\usepackage{amsmath} %for dealing with mathematics,
\usepackage{amsthm} %for dealing with theorem environments,
\usepackage{url} %It provides better support for handling and breaking URLs.
\usepackage{stmaryrd}
\usepackage{tikz}
\usepackage{physics}
\usepackage{bm}
\usepackage{booktabs}
\usepackage{subcaption}
\usepackage{color}

\newcommand{\Slash}[1]{{\ooalign{\hfil/\hfil\crcr$#1$}}}
\numberwithin{equation}{section}

\allowdisplaybreaks

\begin{document}

\title{A basis of the gradient flow exact renormalization group for gauge
theory}

%%%% To generate auto affiliation numbers please use \author{}\affil{} command

\author[1]{Hidenori Sonoda}
\affil[1]{Department of Physics and Astronomy, the University of Iowa, Iowa City, Iowa 52242, USA}

\author[2]{Hiroshi Suzuki}
\affil[2]{Department of Physics, Kyushu University, 744 Motooka, Nishi-ku,
Fukuoka 819-0395, Japan}

%% \author{Insert second author name here}
%% \affil{Insert second author address here}

%% \author{Insert third author name here}
%% \author[3]{Insert fourth author name here} %%% Use optional bracket [3] to change the respective address
%% \affil{Insert third author address here}

%% \author{Insert last author name here\thanks{These authors contributed equally to this work}}
%% \affil{Insert last author address here}

%%% To include the collaborator name... Please use the command "\collaborator"
%%% For example: \collaborator{ATLAS Collaboration}

\begin{abstract}%
The gradient flow exact renormalization group (GFERG) is a variant of the exact
renormalization group (ERG) for gauge theory that is aimed at preserve gauge
invariance as manifestly as possible. It achieves this goal by utilizing the
Yang--Mills gradient flow or diffusion for the block-spin process. In this
paper, we formulate GFERG by the Reuter equation in which the block spinning is
done by Gaussian integration. This formulation provides a simple understanding
of various points of GFERG, unresolved thus far. First, there exists a unique
ordering of functional derivatives in the GFERG equation that remove ambiguity
of contact terms. Second, perturbation theory of GFERG suffers from
unconventional ultraviolet (UV) divergences if no gauge fixing is introduced.
This explains the origin of some UV divergences we have encountered in
perturbative solutions to GFERG. Third, the modified correlation functions
calculated with the Wilson action in GFERG coincide with the correlation
functions of diffused or flowed fields calculated with the bare action. This
shows the existence of a Wilson action that reproduces precisely the physical
quantities computed by the gradient flow formalism (up to contact terms). We
obtain a definite ERG interpretation of the gradient flow. The formulation
given in this paper provides a basis for further perturbative/nonperturbative
computations in GFERG, preserving gauge invariance maximally.
\end{abstract}

\subjectindex{B01,B05,B31,B32}

\maketitle

\section{Introduction}
\label{sec:1}
The exact renormalization group
(ERG)~\cite{Wilson:1973jj,Morris:1993qb,Becchi:1996an,Pawlowski:2005xe,Igarashi:2009tj,Rosten:2010vm,Dupuis:2020fhh}
is a framework that helps us understand possible local field theories in
continuous spacetime. When applied to gauge theory, a momentum cutoff inherent
in the ERG formalism is compatible with gauge invariance, but the invariance is
not manifest, expressed as a relation nonlinear with respect to the Wilson
action. Given a nonperturbative ansatz for the Wilson action, it is difficult
to tell if the action has gauge invariance or not. A manifest gauge symmetry in
ERG would thus be highly desirable.

The gradient flow exact renormalization group (GFERG), considered
in~Refs.~\cite{Sonoda:2020vut,Miyakawa:2021hcx,Miyakawa:2021wus,Sonoda:2022fmk,Miyakawa:2023yob},
is aimed at achieve manifest gauge invariance within the ERG formalism;
see also~Refs.~\cite{Abe:2022smm,Miyakawa:2022qbz,Haruna:2023spq} for
related works.\footnote{See Refs.~\cite{Morris:1999px,Morris:2000fs,Arnone:2005fb,Morris:2005tv} for the works of Morris et~al.\ with the same motivation.} The
idea of GFERG is based on an observation that the block-spin process in ERG for
scalar field theory can be understood in terms of heat
diffusion~\cite{Sonoda:2019ibh}; see
also~Refs.~\cite{Matsumoto:2020lha,Tanaka:2022pwt}. This observation leads to a
precise connection between ERG and the gradient
flow~\cite{Narayanan:2006rf,Luscher:2010iy,Luscher:2011bx,Luscher:2013cpa}: the
modified correlation functions~\cite{Sonoda:2015bla} given by a Wilson action
reproduce the correlation functions of the scalar field~\cite{Capponi:2015ucc}
subject to the gradient flow. Thanks to this connection, we can understand
intuitively that only the standard wave function renormalization of the
elementary scalar field suffices to render the composite operators ultraviolet
(UV) finite in the gradient flow formalism~\cite{Capponi:2015ucc}.

Having understood a connection between ERG and the diffusion equation, it is
natural to formulate ERG for gauge theory by employing the Yang--Mills gradient
flow equation~\cite{Narayanan:2006rf,Luscher:2010iy,Luscher:2011bx} that
preserves gauge covariance. As expected, the resulting ERG can preserve gauge
invariance~\cite{Sonoda:2020vut}. We can incorporate the fermion
field~\cite{Miyakawa:2021hcx,Miyakawa:2021wus} by using the quark
flow~\cite{Luscher:2013cpa}. Some issues were left unresolved by the previous
works, however. Firstly, there was ambiguity on the ordering of fields and
functional derivatives in the GFERG flow equation. Secondly, it has been not
clear whether perturbation theory in GFERG requires gauge fixing or not; actual
perturbative calculations in GFERG~\cite{Miyakawa:2021wus,Sonoda:2025xxx}
indicate that, without gauge fixing, perturbation theory of GFERG suffers from
\emph{unconventional\/} UV divergences. Thirdly, the relation between two
correlation functions, one computed by the Wilson action of GFERG and another
of the diffused or flowed fields obeying the flow
equations~\cite{Luscher:2010iy,Luscher:2011bx,Luscher:2013cpa} has not been
pinned down precisely.

In the present paper, we formulate GFERG in the form of the Reuter
equation\footnote{%
See, for instance, Eqs.~(2.4) and~\eqref{eq:(2.5)}
of~Ref.~\cite{Reuter:2019byg}. Our Eq.~\eqref{eq:(2.3)} is obtained by setting
$W_k[J]\to S_\Lambda[\phi]$, where
$J\to\Lambda Z(\Lambda)e^{-tp^2}\phi(p)$, and
$\Hat{\phi}\to\phi'$
and~$\mathcal{R}_k(p^2)\to\Lambda^2Z(\Lambda)^2e^{-2tp^2}$.} in which the block
spin process is expressed by the Gaussian averaging. This formulation, after
some careful consideration, provides answers to the above three points.
Regarding the first point, there exists a well-defined ordering of fields and
functional derivatives in the GFERG flow equation which is now free from the
ambiguity of contact terms. As for the second, it becomes clear that
perturbation theory of GFERG generally suffers from unconventional UV
divergences if no gauge fixing is introduced. As for the third, it becomes
quite clear that the modified correlation functions
in~Ref.~\cite{Sonoda:2015bla} defined by the Wilson action of GFERG coincide
with the correlation functions of diffused or flowed fields calculated by the
bare action; hence, there exists a Wilson action that reproduces precisely the
physical quantities computed by the gradient flow formalism in gauge theory. We
obtain a definite ERG interpretation of the gradient
flow~\cite{Luscher:2013vga,Kagimura:2015via,Yamamura:2015kva,Aoki:2016ohw,Pawlowski:2017rhn,Makino:2018rys,Abe:2018zdc,Carosso:2018bmz,Carosso:2018rep,Carosso:2019qpb,Morikawa:2024nrp}.

We therefore believe that this paper provides a solid basis for further
perturbative/nonperturbative computations within the GFERG formulation that
preserves gauge invariance maximally.

Let us summarize the content of the paper. In Section~\ref{sec:2}, we review
the interpretation of the gradient flow in scalar field
theory~\cite{Capponi:2015ucc} in terms of an ERG formulation based on a simple
heat diffusion equation~\cite{Sonoda:2019ibh}. This example illustrates the
basic idea and manipulations which will be repeatedly employed in GFERG applied
to gauge theory. In~Section~\ref{sec:3}, we formulate GFERG using the Reuter
equation. We do not assume any gauge fixing so that the GFERG flow equation
preserves the \emph{conventional\/} gauge invariance; a gauge invariant Wilson
action keeps its invariance under the GFERG flow. We believe that this
remarkable feature of GFERG is quite helpful in setting a nonperturbative
ansatz for the Wilson action in gauge theory. On the other hand, from the
integral representation of the Wilson action, we suspect that without gauge
fixing perturbative expansions of GFERG suffer from unconventional UV
divergences. Therefore, in~Sections~\ref{sec:4} and~\ref{sec:5}, we develop
GFERG with gauge fixing. Section~\ref{sec:4} is for a formulation with the
Nakanishi--Lautrup (NL) field, whereas Section~\ref{sec:5} is for a formulation
without the NL field. The underlying BRST (rather than gauge) symmetry is also
preserved under GFERG. Our guiding principle for the formulation is that the
modified correlation functions~\cite{Sonoda:2015bla} defined by the Wilson
action reproduce the correlation functions of diffused or flowed fields defined
in~Ref.~\cite{Luscher:2011bx} (see also~Ref.~\cite{Hieda:2016xpq}). This
requirement leads naturally to a prescription that the Faddeev--Popov (FP)
anti-ghost and the NL fields are not diffused or flowed. In the formulation
of~Section~\ref{sec:4} with the NL field, the Ward--Takahashi (WT) identity
associated with the BRST symmetry takes a relatively simple closed form. In the
formulation of~Section~\ref{sec:5} without the NL field, the WT identity
becomes infinite order in the Wilson action. Still, the dependence on the NL
field can be quite nontrivial in the formulation of~Section~\ref{sec:4}. Thus,
we do not take a side, giving both formulations in the present paper for future
developments. We note that the formulation in~Section~\ref{sec:5}, when applied
to quantum electrodynamics (QED), reproduces precisely the GFERG formulation
given in~Ref.~\cite{Miyakawa:2021wus}. (In Abelian gauge theories, the
WT identity becomes linear in the Wilson action.) In~Section~\ref{sec:6}, for
completeness, we recapitulate GFERG for the one-particle (1PI) Wilson
action~\cite{Morris:1993qb,Nicoll:1977hi,Wetterich:1992yh,Wetterich:1993ne,Bervillier:2008an}
following~Ref.~\cite{Sonoda:2022fmk}. Section~\ref{sec:7} concludes the paper.

\section{Scalar field theory}
\label{sec:2}
\subsection{Wilson ERG and field diffusion}
\label{sec:2.1}
We can use scalar field theory~\cite{Sonoda:2019ibh} to illustrate the idea of
GFERG. Given a bare action~$S[\phi']$, we define the Wilson
action~$S_\Lambda[\phi]$ with a UV cutoff~$\Lambda$ by the Reuter
formula\footnote{%
An equivalent definition is given by
\begin{equation}
   e^{S_\Lambda}[\phi]
   =\Hat{s}\int[d\phi']\,
   \prod_x\delta\left(\phi(x)-Z(\Lambda)\phi'(t,x)\right)
   e^{S[\phi']},
\label{eq:(2.1)}
\end{equation}
where the ``scrambler'' $\Hat{s}$ is the differential operator
\begin{equation}
   \Hat{s}:=
   \exp\left[\frac{1}{2\Lambda^2}\int d^Dx\,
   \frac{\delta^2}{\delta\phi(x)\delta\phi(x)}\right].
\label{eq:(2.2)}
\end{equation}
Some references
in~Refs.~\cite{Sonoda:2020vut,Miyakawa:2021hcx,Miyakawa:2021wus,Sonoda:2022fmk,Miyakawa:2023yob} use this instead of~Eq.~\eqref{eq:(2.3)}.}
\begin{equation}
   e^{S_\Lambda[\phi]}
   :=\mathcal{N}(\Lambda)
   \int[d\phi']\,
   \exp\left\{
   -\frac{\Lambda^2}{2}\int d^Dx\,
   \left[
   \phi(x)-Z(\Lambda)\phi'(t,x)
   \right]^2\right\}e^{S[\phi']},
\label{eq:(2.3)}
\end{equation}
where $D$ denotes the spacetime dimension, the normalization factor is given by
\begin{equation}
   \mathcal{N}(\Lambda):=
   \left\{\int[d\phi]\,
   \exp\left[
   -\frac{\Lambda^2}{2}\int d^Dx\,\phi(x)^2\right]\right\}^{-1},
\label{eq:(2.4)}
\end{equation}
and $Z(\Lambda)$ is a possible wave function renormalization factor. The field
variable~$\phi'(x)$ of the bare action is related to $\phi'(t,x)$ in the
Gaussian factor by the \emph{diffusion equation}
\begin{equation}
   \partial_t\phi'(t,x)=\partial_\mu\partial_\mu\phi'(t,x),\qquad
   \phi'(t=0,x)=\phi'(x),
\label{eq:(2.5)}
\end{equation}
where the diffusion (or flow) time~$t$ and the cutoff~$\Lambda$ are related as
\begin{equation}
   t:=\frac{1}{\Lambda^2}-\frac{1}{\Lambda_0^2}.
\label{eq:(2.6)}
\end{equation}
The momentum $\Lambda_0$ gives a scale where the bare action~$S[\phi']$ is
defined. Note that, as~$\Lambda$ changes from~$\Lambda_0$ to~$0$, $t$ changes
from~$t=0$ to~$\infty$. The idea behind~Eq.~\eqref{eq:(2.3)} is quite natural:
the Wilson action is given by the block-spin transformation defined by the
combination of diffusion~\eqref{eq:(2.5)} and the Gaussian averaging. The
normalization factor~$\mathcal{N}(\Lambda)$ ensures the invariance of the
partition function under the change of~$\Lambda$, i.e.,
$\int[d\phi]\,e^{S_{\Lambda}[\phi]}=\int[d\phi]\,e^{S[\phi]}$.

The ERG equation for~$S_\Lambda[\phi]$ can be obtained by the
$\Lambda$~derivative of~Eq.~\eqref{eq:(2.3)}. Using
\begin{equation}
   \Lambda\frac{dt}{d\Lambda}=-\frac{2}{\Lambda^2}
\label{eq:(2.7)}
\end{equation}
and Eq.~\eqref{eq:(2.5)}, we find
\begin{align}
   &-\Lambda\frac{\partial}{\partial\Lambda}
   e^{S_\Lambda[\phi]}
\notag\\
   &=\mathcal{N}(\Lambda)
   \int[d\phi']\,
   \exp\left\{
   -\frac{\Lambda^2}{2}\int d^Dx\,
   \left[
   \phi(x)-Z(\Lambda)\phi'(t,x)
   \right]^2\right\}e^{S[\phi']}
\notag\\
   &\qquad\qquad{}
   \times\int d^Dx\,
   \biggl\{
   -\delta^D(0)
   +\Lambda^2\left[\phi(x)-Z(\Lambda)\phi'(t,x)\right]^2
\notag\\
   &\qquad\qquad\qquad\qquad\qquad{}
   +\Lambda^2\left[\phi(x)-Z(\Lambda)\phi'(t,x)\right]
   \left[
   \frac{2}{\Lambda^2}\partial_\mu\partial_\mu
   -\Lambda\frac{d\ln Z(\Lambda)}{d\Lambda}\right]
   Z(\Lambda)\phi'(t,x)\biggr\}.
\label{eq:(2.8)}
\end{align}
We can replace $\phi'(t,x)$ on the right-hand side by the functional derivative
of the Gaussian factor in the integrand of~Eq.~\eqref{eq:(2.3)}. The dictionary
for the replacement is given by
\begin{equation}
   \phi(x)-Z(\Lambda)\phi'(t,x)\leftrightarrow
   -\frac{1}{\Lambda^2}\frac{\delta}{\delta\phi(x)},\qquad
   Z(\Lambda)\phi'(t,x)\leftrightarrow
   \phi(x)+\frac{1}{\Lambda^2}\frac{\delta}{\delta\phi(x)}.
\label{eq:(2.9)}
\end{equation}
For the second order derivative, we have to be careful about the ordering of
fields and functional derivatives. We find
\begin{align}
   &-\Lambda\frac{\partial}{\partial\Lambda}
   e^{S_\Lambda[\phi]}
\notag\\
   &=\int d^Dx\,
   \frac{\delta}{\delta\phi(x)}\left(
   \left\{
   -\left[
   \frac{2}{\Lambda^2}\partial_\mu\partial_\mu
   +\gamma(\Lambda)
   \right]
   \left[
   \phi(x)+\frac{1}{\Lambda^2}\frac{\delta}{\delta\phi(x)}\right]
   +\frac{1}{\Lambda^2}\frac{\delta}{\delta\phi(x)}
   \right\}
   e^{S_\Lambda[\phi]}\right),
\label{eq:(2.10)}
\end{align}
where we have introduced the anomalous dimension by
\begin{equation}
   \gamma(\Lambda):=-\Lambda\frac{d\ln Z(\Lambda)}{d\Lambda}.
\label{eq:(2.11)}
\end{equation}
Note that the right-hand side of~Eq.~\eqref{eq:(2.10)} is a total functional
derivative reflecting the invariance of the partition function
$-\Lambda d/d\Lambda\int[d\phi]\,e^{S_{\Lambda}[\phi]}=0$.
Equation~\eqref{eq:(2.10)} is nothing but Wilson's ERG equation for scalar
field theory; in the notation of~Ref.~\cite{Sonoda:2015bla}, cutoff functions
that parametrize ERG equations are given by~$K(p/\Lambda)=e^{-p^2/\Lambda^2}$
and~$k(p/\Lambda)=p^2/\Lambda^2$ in the momentum space.

We now define a generating functional of correlation functions associated with
the Wilson action~$S_\Lambda$ by
\begin{align}
   \mathcal{Z}[J]&:=\int[d\phi]\,e^{S_\Lambda[\phi]}
   \exp\left[-\frac{1}{2\Lambda^2}\int d^Dx\,
   \frac{\delta^2}{\delta\phi(x)\delta\phi(x)}\right]
   e^{\int d^Dx\,J(x)\phi(x)}
\notag\\
   &=\exp\left[-\frac{1}{2\Lambda^2}\int d^Dx\,J(x)J(x)\right]
   \int[d\phi]\,e^{S_\Lambda[\phi]}
   e^{\int d^Dx\,J(x)\phi(x)}.
\label{eq:(2.12)}
\end{align}
The correlation function given by the functional derivatives of
this~$\mathcal{Z}[J]$ is the ``modified'' correlation function
introduced in~Ref.~\cite{Sonoda:2015bla} and differs from the conventional
correlation function by contact terms. The modified correlation function
exhibits a simpler scaling property under the scale
transformation~\cite{Sonoda:2015bla}. Substituting Eq.~\eqref{eq:(2.3)} into
this and integrating over~$\phi$, we obtain
\begin{align}
   \mathcal{Z}[J]&=\exp\left[-\frac{1}{2\Lambda^2}\int d^Dx\,J(x)J(x)\right]
   \mathcal{N}(\Lambda)
\notag\\
   &\qquad{}
   \times
   \int[d\phi']\,e^{S[\phi']}
   \int[d\phi]\,
   \exp\left\{
   -\frac{\Lambda^2}{2}\int d^Dx\,
   \left[\phi(x)-Z(\Lambda)\phi'(t,x)\right]^2\right\}
   e^{\int d^Dx\,J(x)\phi(x)}
\notag\\
   &=\int[d\phi']\,e^{S[\phi']}e^{\int d^Dx\,J(x)Z(\Lambda)\phi'(t,x)}.
\label{eq:(2.13)}
\end{align}
The equality between~Eqs.~\eqref{eq:(2.12)} and~\eqref{eq:(2.13)} implies that
\begin{align}
   \left\langle
   \exp\left[-\frac{1}{2\Lambda^2}\int d^Dx\,
   \frac{\delta^2}{\delta\phi(x)\delta\phi(x)}\right]
   \phi(x_1)\dotsb\phi(x_N)
   \right\rangle_{S_\Lambda}
   =Z(\Lambda)^N\left\langle\phi(t,x_1)\dotsb\phi(t,x_N)
   \right\rangle_S,
\label{eq:(2.14)}
\end{align}
where the subscripts $S_\Lambda$ and~$S$ indicate that the correlation functions
are computed by employing the actions $S_\Lambda$ and~$S$, respectively. Hence,
the modified correlation function computed by the Wilson action~$S_\Lambda$ with
the cutoff~$\Lambda$ reproduces the correlation function of the diffused field
obeying~Eq.~\eqref{eq:(2.5)}, computed by the bare action~$S$ with the
cutoff~$\Lambda_0$~\cite{Sonoda:2019ibh}. Equation~\eqref{eq:(2.14)} thus
provides a precise connection between the Wilson ERG and the gradient
flow~\cite{Narayanan:2006rf,Luscher:2010iy} in scalar field
theory~\cite{Capponi:2015ucc}. Suppose that we renormalize the bare action~$S$:
we tune the coupling constants and the wave function renormalization
factor~$Z(\Lambda)$ appropriately so that the correlation function on the
right-hand side of~Eq.~\eqref{eq:(2.14)} is finite as we
take~$\Lambda_0\to\infty$. The Wilson action~$S_\Lambda$ for a fixed $\Lambda$
stays finite in this limit. We then expect that the correlation
function~\eqref{eq:(2.14)} is finite even at the equal point
limit~$x_1\to x_2=x$ because the momentum modes of~$S_\Lambda$ are effectively
limited by the cutoff~$\Lambda$. In this way we can understand quite simply why
simple products of the renormalized diffused field, such as
$Z(\Lambda)^2\phi(t,x)^2$, are finite under the conventional
renormalization~\cite{Capponi:2015ucc} at least for an equal diffusion
time~$t$~\cite{Sonoda:2019ibh}.

\subsection{Dimensionless formulation}
\label{sec:2.2}
The Wilson ERG equation~\eqref{eq:(2.10)} describes how the Wilson action
changes as the UV cutoff~$\Lambda$ changes. We sometimes want an equation that
describes how the Wilson action changes under the scale transformation on field
variables under a fixed UV cutoff.

This can be done by measuring all quantities using the UV cutoff~$\Lambda$ as
units; with these dimensionless variables, the UV cutoff itself becomes unity,
and the change in~$\Lambda$ becomes the change in momenta or coordinates.

Thus, we introduce
\begin{equation}
   \Bar{\phi}(\Bar{x})
   :=\Lambda^{-d_\phi}\phi(x),\qquad\Bar{x}:=\Lambda x,
\label{eq:(2.15)}
\end{equation}
where $d_\phi$ is the canonical mass dimension of the field~$\phi$
\begin{equation}
   d_\phi=\frac{D-2}{2},
\label{eq:(2.16)}
\end{equation}
and
\begin{equation}
   \Bar{S}_\tau[\Bar{\phi}]:=S_\Lambda[\phi],\qquad
   e^{-\tau}:=\frac{\Lambda}{\mu},
\label{eq:(2.17)}
\end{equation}
where $\mu$ is an arbitrary mass scale. Since $\phi(x)$ in~$S_\Lambda[\phi]$ is
substituted by~$\Lambda^{d_\phi}\Bar{\phi}(\Lambda x)$, and
\begin{equation}
   \frac{\partial}{\partial\tau}
   \left[\Lambda^{d_\phi}\Bar{\phi}(\Lambda x)\right]
   =\Lambda^{d_\phi}
   \left(-d_\phi-\Bar{x}\cdot\frac{\partial}{\partial\Bar{x}}\right)
   \Bar{\phi}(\Bar{x}),
\label{eq:(2.18)}
\end{equation}
we find
\begin{equation}
   \frac{\partial}{\partial\tau}e^{\Bar{S}_\tau[\Bar{\phi}]}
   =-\Lambda\frac{\partial}{\partial\Lambda}e^{S_\Lambda[\phi]}
   +\int d^Dx\,
   \left(-d_\phi-\Bar{x}\cdot\frac{\partial}{\partial\Bar{x}}\right)
   \Bar{\phi}(\Bar{x})\cdot\frac{\delta}{\delta\Bar{\phi}(\Bar{x})}
   e^{\Bar{S}_\tau[\Bar{\phi}]}.
\label{eq:(2.19)}
\end{equation}
The last term is added to the Wilson ERG
equation~\eqref{eq:(2.10)}.\footnote{Strictly speaking, we must also add
$D\Bar{S}_\tau[0]$ since the vacuum energy density scales as~$e^{D\tau}$.}

When we use the field variable in momentum space, the dimensionless variables
are defined by\footnote{%
We will use the abbreviations
\begin{equation}
   \int_p:=\int\frac{d^Dp}{(2\pi)^D}
\label{eq:(2.20)}
\end{equation}
and
\begin{equation}
   \delta(p):=(2\pi)^D\delta^D(p).
\label{eq:(2.21)}
\end{equation}
}
\begin{equation}
   \Bar{\phi}(\Bar{p})
   :=\Lambda^{-d_\phi+D}\phi(p),\qquad\Bar{p}:=p/\Lambda.
\label{eq:(2.22)}
\end{equation}
The relation corresponding to~Eq.~\eqref{eq:(2.19)} is given by
\begin{equation}
   \frac{\partial}{\partial\tau}e^{\Bar{S}_\tau[\Bar{\phi}]}
   =-\Lambda\frac{\partial}{\partial\Lambda}e^{S_\Lambda[\phi]}
   +\int_{\Bar{p}}
   \left(-d_\phi+D+\Bar{p}\cdot\frac{\partial}{\partial\Bar{p}}\right)
   \Bar{\phi}(\Bar{p})\cdot\frac{\delta}{\delta\Bar{\phi}(\Bar{p})}
   e^{\Bar{S}_\tau[\Bar{\phi}]}.
\label{eq:(2.23)}
\end{equation}

\section{Gauge invariant GFERG}
\label{sec:3}
\subsection{Integral representation}
\label{sec:3.1}
In the previous section we have observed a direct connection between the Wilson
ERG and the gradient flow in scalar field theory. In this section, we construct
an ERG for vector-like gauge theory by generalizing this connection in a
natural manner; here for the diffusion of fields we employ the Yang--Mills
gradient flow in gauge theory~\cite{Luscher:2010iy,Luscher:2013cpa}.
Consequently, this ERG, coined as GFERG, preserves manifest gauge invariance
under the ERG flow~\cite{Sonoda:2020vut,Miyakawa:2021hcx}.

As a natural generalization of~Eq.~\eqref{eq:(2.3)} to gauge theory, we set
\begin{align}
   &e^{S_\Lambda[A,\Bar{\psi},\psi]}
\notag\\
   &:=\mathcal{N}(\Lambda)
   \int[dA'][d\psi'][d\Bar{\psi}']\,
\notag\\
   &\qquad{}
   \times\exp\left\{
   -\frac{\Lambda^{D-2}}{2}\int d^Dx\,
   \left[
   A_\mu^a(x)-A_\mu^{\prime a}(t,x)
   \right]^2\right\}
\notag\\
   &\qquad{}
   \times\exp\left\{
   i\Lambda\int d^Dx\,
   \left[\Bar{\psi}(x)-Z_\psi(\Lambda)\Bar{\psi}'(t,x)\right]
   \left[\psi(x)-Z_\psi(\Lambda)\psi'(t,x)\right]
   \right\}e^{S[A',\Bar{\psi}',\psi']},
\label{eq:(3.1)}
\end{align}
where
\begin{align}
   \mathcal{N}(\Lambda):=
   \left\{\int[dA][d\psi][d\Bar{\psi}]\,
   \exp\left[
   -\frac{\Lambda^{D-2}}{2}\int d^Dx\,A_\mu^a(x)^2\right]
   \exp\left[
   i\Lambda\int d^Dx\,\Bar{\psi}(x)\psi(x)\right]\right\}^{-1}.
\label{eq:(3.2)}
\end{align}
The diffusion equation in scalar field theory, Eq.~\eqref{eq:(2.5)}, is
replaced by the gauge covariant diffusion equation
in~Refs.~\cite{Luscher:2010iy,Luscher:2013cpa}. For the gauge field, it is
given by
\begin{equation}
   \partial_tA_\mu'(t,x)=D_\nu'F_{\nu\mu}'(t,x),\qquad
   A_\mu'(t=0,x)=A_\mu'(x),
\label{eq:(3.3)}
\end{equation}
where
\begin{equation}
   F_{\mu\nu}'(t,x):=\partial_\mu A_\nu'(t,x)-\partial_\nu A_\mu'(t,x)
   +[A_\mu'(t,x),A_\nu'(t,x)],\quad
   D_\mu':=\partial_\mu+[A_\mu'(t,x),\phantom{X}],
\label{eq:(3.4)}
\end{equation}
and for the fermion fields,
\begin{align}
   \partial_t\psi'(t,x)
   &=D_\mu'D_\mu'\psi'(t,x),&
   \psi'(t=0,x)&=\psi'(x),
\notag\\
   \partial_t\Bar{\psi}'(t,x)
   &=\Bar{\psi}'(t,x)
   \overleftarrow{D}_\mu'\overleftarrow{D}_\mu',&
   \Bar{\psi}'(t=0,x)&=\Bar{\psi}'(x),   
\label{eq:(3.5)}
\end{align}
where the covariant derivatives
\begin{equation}
   D_\mu':=\partial_\mu+A_\mu'(t,x),\qquad
   \overleftarrow{D}_\mu':=\overleftarrow{\partial}_\mu-A_\mu'(t,x),
\label{eq:(3.6)}
\end{equation}
are given with $A_\mu'$ in the appropriate representation for the fermion
fields. We note that, in~Eq.~\eqref{eq:(3.1)}, the wave function
renormalization factor for the gauge potential is set to unity,
$Z_A(\Lambda)=1$, since the diffused or flowed gauge potential does not require
any wave function renormalization in the gradient flow
formalism~\cite{Luscher:2010iy,Luscher:2011bx,Luscher:2013cpa}.

\subsection{Gauge invariance of the Wilson action}
\label{sec:3.2}
Let us first show the gauge invariance of the Wilson action~\eqref{eq:(3.1)},
assuming that the bare action~$S$ is gauge invariant.\footnote{%
We discuss whether the integral representation~\eqref{eq:(3.1)} converges or
not in~Section~\ref{sec:3.6}.} In~Eq.~\eqref{eq:(3.1)}, consider the change of
integration variables in the form of an infinitesimal gauge transformation:
\begin{equation}
   A_\mu'(x)\to A_\mu'(x)+D_\mu'\omega(x),\qquad
   \psi'(x)\to\left[1-\omega(x)\right]\psi'(x),\qquad
   \Bar{\psi}'(x)\to\Bar{\psi}'(x)\left[1+\omega(x)\right].
\label{eq:(3.7)}
\end{equation}
The bare action~$S$ does not change under these. Because of the gauge
covariance of the diffusion equations in~Eqs.~\eqref{eq:(3.3)}
and~\eqref{eq:(3.5)}, the gauge variations~\eqref{eq:(3.7)} on the initial
values induce those for the solutions to the diffusion equations.
Under~Eq.~\eqref{eq:(3.7)}, we find
\begin{align}
   A_\mu'(t,x)&\to A_\mu'(t,x)+D_\mu'\omega(x),&&
\notag\\
   \psi'(t,x)&\to\left[1-\omega(x)\right]\psi'(t,x),&
   \Bar{\psi}'(t,x)&\to\Bar{\psi}'(t,x)\left[1+\omega(x)\right].
\label{eq:(3.8)}
\end{align}
Therefore, the quadratic forms in the Gaussian factor in~Eq.~\eqref{eq:(3.1)}
change under~Eq.~\eqref{eq:(3.7)} as\footnote{%
$f^{abc}$ is the structure constant defined by~$[T^a,T^b]=f^{abc}T^c$.}
\begin{align}
   &\left[A_\mu^a(x)-A_\mu^{\prime a}(t,x)\right]^2
\notag\\
   &\to
   \left\{A_\mu^a(x)
   -\left[
   A_\mu^{\prime a}(t,x)+\partial_\mu\omega^a(x)
   +f^{abc}A_\mu^{\prime b}(t,x)\omega^c(x)
   \right]
   \right\}^2
\notag\\
   &=
   %\left[\delta^{ab}-f^{abc}\omega^c(x)\right]
   \left[A_\mu^a(x)-\partial_\mu\omega^a(x)-f^{abc}A_\mu^b(x)\omega^c(x)
   -A_\mu^{\prime a}(t,x)\right]^2
\label{eq:(3.9)}
\end{align}
and
\begin{align}
   &\left[\Bar{\psi}(x)-Z_\psi(\Lambda)\Bar{\psi}'(t,x)\right]
   \left[\psi(x)-Z_\psi(\Lambda)\psi'(t,x)\right]
\notag\\
   &\to
   \left\{
   \Bar{\psi}(x)-Z_\psi(\Lambda)\Bar{\psi}'(t,x)\left[1+\omega(x)\right]
   \right\}\left\{
   \psi(x)-Z_\psi(\Lambda)\left[1-\omega(x)\right]\psi'(t,x)
   \right\}
\notag\\
   &=\left\{\Bar{\psi}(x)\left[1-\omega(x)\right]-Z_\psi(\Lambda)
   \Bar{\psi}'(t,x)\right\}
   \left\{\left[1+\omega(x)\right]\psi(x)-Z_\psi(\Lambda)\psi'(t,x)\right\}.
\label{eq:(3.10)}
\end{align}
Since the functional integral~\eqref{eq:(3.1)} does not change under the change
of integration variables in~Eq.~\eqref{eq:(3.1)},\footnote{%
We assume that the measure of functional integration is gauge invariant.} we
obtain the WT identity\footnote{%
Unless otherwise noted, the functional derivative by a Grassmann-odd variable
is understood as a left-derivative.}
\begin{equation}
   \int d^Dx\,
   \left[
   D_\mu\omega^a(x)\frac{\delta}{\delta A_\mu^a(x)}
   -\omega(x)\psi(x)\frac{\delta}{\delta \psi(x)}
   +\Bar{\psi}(x)\omega(x)\frac{\delta}{\delta\Bar{\psi}(x)}
   \right]e^{S_\Lambda[A,\Bar{\psi},\psi]}=0.
\label{eq:(3.11)}
\end{equation}
This shows that the Wilson
action~$S_\Lambda[A,\Bar{\psi},\psi]$~\eqref{eq:(3.1)} is invariant
under the \emph{conventional\/} gauge transformation if the bare action~$S$ is.
This is remarkable since the ordinary Wilson ERG on the basis of a momentum
cutoff cannot preserve manifest gauge invariance. Note that the above argument
does not hold unless the wave function renormalization factor for the gauge
field~$Z_A(\Lambda)$ is set to unity, $Z_A(\Lambda) = 1$. The gauge invariance
of the Wilson action~\eqref{eq:(3.1)} is strongly desirable in setting up a
nonperturbative ansatz for the Wilson action that is consistent with gauge
invariance.

\subsection{GFERG equation}
\label{sec:3.3}
The ERG equation that the Wilson action~\eqref{eq:(3.1)} obeys can be readily
obtained by taking the $\Lambda$ derivative of~Eq.~\eqref{eq:(3.1)} as
follows:\footnote{%
$\tr_R$ denotes a trace in the gauge representation to which the fermion fields
belong.}
\begin{align}
   &-\Lambda\frac{\partial}{\partial\Lambda}
   e^{S_\Lambda[A,\Bar{\psi},\psi]}
\notag\\
   &=\mathcal{N}(\Lambda)
   \int[dA'][d\psi'][d\Bar{\psi}']\,
\notag\\
   &\qquad{}
   \times\exp\left\{
   -\frac{\Lambda^{D-2}}{2}\int d^Dx\,
   \left[
   A_\mu^a(x)-A_\mu^{\prime a}(t,x)
   \right]^2\right\}
\notag\\
   &\qquad{}
   \times\exp\left\{
   i\Lambda\int d^Dx\,
   \left[\Bar{\psi}(x)-Z_\psi(\Lambda)\Bar{\psi}'(t,x)\right]
   \left[\psi(x)-Z_\psi(\Lambda)\psi'(t,x)\right]
   \right\}e^{S[A',\Bar{\psi}',\psi']}
\notag\\
   &\qquad{}
   \times
   \int d^Dx\,
   \Bigl\{
   -[(D-2)/2]\delta^{aa}D\delta^D(0)
   +[(D-2)/2]\Lambda^{D-2}\left[A_\mu^a(x)-A_\mu^{\prime a}(t,x)\right]^2
\notag\\
   &\qquad\qquad\qquad\qquad{}
   +2\Lambda^{D-4}\left[A_\mu^a(x)-A_\mu^{\prime a}(t,x)\right]
   D_\nu'F_{\nu\mu}^{\prime a}(t,x)
\notag\\
   &\qquad\qquad\qquad\qquad{}
   +(\tr_R1)2^{[D/2]}\delta^D(0)
   -i\Lambda\left[\Bar{\psi}(x)-Z_\psi\Bar{\psi}'(t,x)\right]
   \left[\psi(x)-Z_\psi\psi'(t,x)\right]
\notag\\
   &\qquad\qquad\qquad\qquad{}
   -\frac{2i}{\Lambda}Z_\psi\Bar{\psi}'(t,x)
   \left[\overleftarrow{D}_\mu'\overleftarrow{D}_\mu'
   +\frac{\Lambda^2}{2}\gamma_\psi(\Lambda)\right]
   \left[\psi(x)-Z_\psi\psi'(t,x)\right]
\notag\\
   &\qquad\qquad\qquad\qquad{}
   -\frac{2i}{\Lambda}
   \left[\Bar{\psi}(x)-Z_\psi\Bar{\psi}'(t,x)\right]
   \left[D_\mu'D_\mu'
   +\frac{\Lambda^2}{2}\gamma_\psi(\Lambda)\right]Z_\psi\psi'(t,x)
   \Bigr\},
\label{eq:(3.12)}
\end{align}
where we have set the anomalous dimension of the fermion field by
\begin{equation}
   \gamma_\psi(\Lambda):=-\Lambda\frac{d\ln Z_\psi(\Lambda)}{d\Lambda}.
\label{eq:(3.13)}
\end{equation}
Note that we have used diffusion equations Eqs.~\eqref{eq:(3.3)}
and~\eqref{eq:(3.5)} here. Then, as Eq.~\eqref{eq:(2.10)}, the right-hand side
of~Eq.~\eqref{eq:(3.12)} can be expressed in terms of the Wilson action as
\begin{align}
   &-\Lambda\frac{\partial}{\partial\Lambda}
   e^{S_\Lambda[A,\Bar{\psi},\psi]}
\notag\\
   &=\int d^Dx\,
   \biggl(
   \frac{\delta}{\delta A_\mu^a(x)} 
   \left\{
   -\frac{2}{\Lambda^2}
   \Hat{D}_\nu\Hat{F}_{\nu\mu}^a(x)
   +[(D-2)/2]\frac{1}{\Lambda^{D-2}}
   \frac{\delta}{\delta A_\mu^a(x)}\right\}
\notag\\
   &\qquad\qquad\qquad{}
   +\frac{\delta}{\delta\Bar{\psi}(x)}\cdot\left\{
   \Hat{\Bar{\psi}}(x)
   \left[
   \frac{2}{\Lambda^2}
   \Hat{\overleftarrow{D}}_\mu\Hat{\overleftarrow{D}}_\mu
   +\gamma_\psi(\Lambda)\right] 
   \right\}
\notag\\
   &\qquad\qquad\qquad{}
   +\frac{\delta}{\delta\psi(x)}\left\{
   \left[
   \frac{2}{\Lambda^2}
   \Hat{D}_\mu\Hat{D}_\mu
   +\gamma_\psi(\Lambda)\right]
   \Hat{\psi}(x)\right\}
\notag\\
   &\qquad\qquad\qquad{}
   -\frac{i}{\Lambda}\frac{\delta}{\delta\psi(x)}
   \frac{\delta}{\delta\Bar{\psi}(x)}\biggr)
   e^{S_\Lambda[A,\Bar{\psi},\psi]},
\label{eq:(3.14)}
\end{align}
where we have carefully chosen the order of functional derivatives so that it
reproduces~Eq.~\eqref{eq:(3.12)} without any contact terms. In this expression,
it is understood that fields in the hatted quantities, such as
$\Hat{F}_{\nu\mu}^a$ and~$\Hat{D}_\nu$, are replaced by
\begin{align}
   \Hat{A}_\mu^a(x)
   &:=A_\mu^a(x)+\frac{1}{\Lambda^{D-2}}\frac{\delta}{\delta A_\mu^a(x)},&&
\notag\\
   \Hat{\psi}(x)
   &:=\psi(x)+\frac{i}{\Lambda}\frac{\delta}{\delta\Bar{\psi}(x)},&
   \Hat{\Bar{\psi}}(x)
   &:=\Bar{\psi}(x)-\frac{i}{\Lambda}\frac{\delta}{\delta\psi(x)}.
\label{eq:(3.15)}
\end{align}

In the previous subsection, we have shown the gauge invariance of the Wilson
action~$S_\Lambda$ for any $\Lambda$ as long as the bare action~$S$ is gauge
invariant. This implies that the GFERG flow~\eqref{eq:(3.14)} preserves gauge
invariance. Though the ordinary ERG equation contains at most second order
derivatives in field variables, the GFERG equation~\eqref{eq:(3.14)} contains
up to the fourth order in gauge fields. This complication is the price to pay
for the manifest gauge invariance.

\subsection{Correlation functions}
\label{sec:3.4}
Next, we write down a relation analogous to~Eq.~\eqref{eq:(2.14)}. There is a
subtlety, however, because the computation of correlation functions using a
gauge invariant action with noncompact gauge fields suffers from the
well-known infinite gauge volume problem. Disregarding this subtlety, as we can
show below, the modified correlation functions computed by the gauge invariant
Wilson action~$S_\Lambda$ reproduce, up to contact terms, the correlation
functions of diffused fields computed by, say, lattice gauge theory. Here, of
course, only the correlation functions of gauge invariant operators can be
considered.

Thus, as a natural counterpart of~Eq.~\eqref{eq:(2.12)}, we define the
generating functional of the modified correlation functions as
\begin{align}
   &\mathcal{Z}[J,\eta,\Bar{\eta}]
\notag\\
   &:=
   \int[dA][d\psi][d\Bar{\psi}]\,
   e^{S_\Lambda[A,\Bar{\psi},\psi]}
\notag\\
   &\qquad\qquad{}
   \times\exp\left[
   -\frac{1}{2\Lambda^{D-2}}\int d^Dx\,
   \frac{\delta^2}{\delta A_\mu^a(x)\delta A_\mu^a(x)}\right]
   \exp\left[
   \frac{i}{\Lambda}\int d^Dx\,
   \frac{\delta}{\delta\psi(x)}\frac{\delta}{\delta\Bar{\psi}(x)}\right]
\notag\\
   &\qquad\qquad\qquad{}
   \times\exp\left\{
   \int d^Dx\,J_\mu^a(x)A_\mu^a(x)
   +\int d^Dx\,\left[
   \Bar{\eta}(x)\psi(x)+\Bar{\psi}(x)\eta(x)
   \right]\right\}.
\label{eq:(3.16)}
\end{align}
Then, using~Eq.~\eqref{eq:(3.1)} and following the same procedure as
in~Eqs.~\eqref{eq:(2.12)} and~\eqref{eq:(2.13)}, we find
\begin{align}
   &\mathcal{Z}[J,\eta,\Bar{\eta}]
\notag\\
   &=
   \int[dA'][d\psi'][d\Bar{\psi}']\,
   e^{S[A',\Bar{\psi}',\psi']}
\notag\\
   &\qquad{}
   \times\exp\left\{
   \int d^Dx\,J_\mu^a(x)A_\mu^{\prime a}(t,x)
   +\int d^Dx\,\left[
   \Bar{\eta}(x)Z_\psi(\Lambda)\psi'(t,x)
   +Z_\psi(\Lambda)\Bar{\psi}'(t,x)\eta(x)
   \right]\right\}.
\label{eq:(3.17)}
\end{align}
The equality of Eqs.~\eqref{eq:(3.16)} and~\eqref{eq:(3.17)} implies that
\begin{align}
   &\biggl\langle
   \exp\left[
   -\frac{1}{2\Lambda^{D-2}}\int d^Dx\,
   \frac{\delta^2}{\delta A_\mu^a(x)\delta A_\mu^a(x)}\right]
   \exp\left[
   \frac{i}{\Lambda}\int d^Dx\,
   \frac{\delta}{\delta\psi(x)}\frac{\delta}{\delta\Bar{\psi}(x)}\right]
\notag\\
   &\qquad\qquad\qquad{}
   \times
   A_{\mu_1}^{a_1}(x_1)\dotsb A_{\mu_N}^{a_N}(x_N)
   \psi(y_1)\dotsb\psi(y_M)\Bar{\psi}(z_1)\dotsb\Bar{\psi}(z_L)
   \biggr\rangle_{S_\Lambda}
\notag\\
   &=Z_\psi(\Lambda)^{M+L}\biggl\langle
   A_{\mu_1}^{a_1}(t,x_1)\dotsb A_{\mu_N}^{a_N}(t,x_N)
   \psi(t,y_1)\dotsb\psi(t,y_M)\Bar{\psi}(t,z_1)\dotsb\Bar{\psi}(t,z_L)
   \biggr\rangle_S,
\label{eq:(3.18)}
\end{align}
which is analogous to~Eq.~\eqref{eq:(2.14)}. Here, we have a tacit
understanding that gauge invariant combinations are taken. It is remarkable
that the modified correlation functions computed by the Wilson
action~$S_\Lambda$ are identical, up to the wave function renormalization
factor~$Z_\psi(\Lambda)$ and contact terms, to the correlation functions of the
diffused fields computed by the bare action~$S$.\footnote{%
Our $A_\mu(t,x)$, $\psi(t,x)$ and~$\Bar{\psi}(t,x)$ are $B_\mu(t,x)$,
$\chi(t,x)$, and~$\Bar{\chi}(t,x)$
in~Refs.~\cite{Luscher:2010iy,Luscher:2011bx,Luscher:2013cpa}, respectively.}
For this equivalence, the fields are constrained to have an equal diffusion
time~$t$, which corresponds to the momentum cutoff~$\Lambda$ in the Wilson
action by~Eq.~\eqref{eq:(2.6)}

This relation provides a connection between a particular form of the Wilson
ERG, i.e., our GFERG, and the gradient flow
formalism~\cite{Luscher:2010iy,Luscher:2011bx,Luscher:2013cpa}; a possible
connection between these has been long anticipated from physical
grounds~\cite{Luscher:2013vga,Kagimura:2015via,Yamamura:2015kva,Aoki:2016ohw,Pawlowski:2017rhn,Makino:2018rys,Abe:2018zdc,Carosso:2018bmz,Carosso:2018rep,Carosso:2019qpb,Morikawa:2024nrp}, and Eq.~\eqref{eq:(3.18)} provides a precise
connection. Recall our observation in~Section~\ref{sec:3.2} that the absence of
wave function renormalization for the gauge field is necessary for the gauge
invariance of the Wilson action. This is consistent with the absence of wave
function renormalization for the diffused gauge field in the gradient flow
formalism~\cite{Luscher:2011bx}.

\subsection{Chiral symmetry}
\label{sec:3.5}
We may consider the chiral symmetry of the Wilson action defined
by~Eq.~\eqref{eq:(3.1)}. Consider the following infinitesimal change of
integration variables:
\begin{equation}
   \psi'(x)\to\left(1+i\theta\gamma_5\right)\psi'(x),\qquad
   \Bar{\psi}'(x)\to\Bar{\psi}'(x)\left(1+i\theta\gamma_5\right),
\label{eq:(3.19)}
\end{equation}
where $\theta$ is an infinitesimal parameter. Assuming that the bare action is
chiral symmetric, i.e., $S[A',\Bar{\psi}',\psi']$ is invariant under this
substitution, and noting that the chiral transformation~\eqref{eq:(3.19)}
induces the same chiral transformation on the diffused fields $\psi'(t,x)$
and~$\Bar{\psi}'(t,x)$, we obtain the identity
\begin{align}
   0&=\int[dA'][d\psi'][d\Bar{\psi}']\,
   \exp\left\{
   -\frac{\Lambda^{D-2}}{2}\int d^Dx\,
   \left[
   A_\mu^a(x)-A_\mu^{\prime a}(t,x)
   \right]^2\right\}
\notag\\
   &\qquad{}
   \times
   \exp\left\{
   i\Lambda\int d^Dx\,
   \left[\Bar{\psi}(x)-Z_\psi(\Lambda)\Bar{\psi}'(t,x)\right]
   \left[\psi(x)-Z_\psi(\Lambda)\psi'(t,x)\right]
   \right\}e^{S[A',\Bar{\psi}',\psi']}
\notag\\
   &\qquad{}
   \times
   \int d^Dx\,
   \bigl\{
   Z_\psi(\Lambda)\Bar{\psi}'(t,x)\gamma_5
   \left[\psi(x)-Z_\psi(\Lambda)\psi'(t,x)\right]
\notag\\
   &\qquad\qquad\qquad\qquad{}
   +\left[\Bar{\psi}(x)-Z_\psi(\Lambda)\Bar{\psi}'(t,x)\right]
   Z_\psi(\Lambda)\gamma_5\psi'(t,x)
   \bigr\},
\label{eq:(3.20)}
\end{align}
This can be expressed as
\begin{equation}
   \int d^Dx\,\left\{\tr
   \frac{\delta}{\delta\psi(x)}
   \left[\gamma_5\Hat{\psi}(x)e^{S_\Lambda[A,\Bar{\psi},\psi]}\right]
   +\tr\frac{\delta}{\delta\Bar{\psi}(x)}\left[\Hat{\Bar{\psi}}(x)
   \gamma_5e^{S_\Lambda[A,\Bar{\psi},\psi]}
   \right]\right\}=0,
\label{eq:(3.21)}
\end{equation}
or
\begin{align}
   &\int d^Dx\,\Biggl\{
   S_\Lambda\frac{\overleftarrow{\delta}}{\delta\psi(x)}
   \gamma_5
   \psi(x)
   +\Bar{\psi}(x)\gamma_5\frac{\delta}{\delta\Bar{\psi}(x)}S_\Lambda
   +\frac{2i}{\Lambda}S_\Lambda\frac{\overleftarrow{\delta}}{\delta\psi(x)}
   \gamma_5\frac{\delta}{\delta\Bar{\psi}(x)} S_\Lambda
\notag\\
   &\qquad\qquad{}
   -\frac{2i}{\Lambda}
   \tr\left[
   \gamma_5
   \frac{\delta}{\delta\Bar{\psi}(x)}
   S_\Lambda
   \frac{\overleftarrow{\delta}}{\delta\psi(x)}\right]
   \Biggr\}=0.
\label{eq:(3.22)}
\end{align}
If the Wilson action is bi-linear in the fermion fields, this reduces to
nothing but the Ginsparg--Wilson relation~\cite{Ginsparg:1981bj}.

\subsection{Finiteness of the perturbative solution to the GFERG equation}
\label{sec:3.6}
So far, we have assumed that the integral representation~\eqref{eq:(3.1)} for
the gauge invariant Wilson action~$S_\Lambda$ is well defined. In the integral
representation, we assume a gauge invariant bare action~$S$ that does not
contain any gauge fixing term and that is regularized gauge invariantly. In
general, without the gauge fixing, perturbation theory for gauge theory is
ill-defined because the propagator does not exist for the longitudinal modes.
Since the integral~\eqref{eq:(3.1)} contains the Gaussian damping factors even
for the longitudinal modes, one might expect that the integral~\eqref{eq:(3.1)}
is well-defined even in perturbation theory. We can see, however, that this
expectation is not necessarily warranted.

To see this, consider a bare action~$S$ given by
\begin{align}
   &S[A',\psi',\Bar{\psi}']
\notag\\
   &=-\frac{1}{g_0^2}\int d^Dx\,\left[
   \frac{1}{4}F_{\mu\nu}^{\prime a}(x)F_{\mu\nu}^{\prime a}(x)
   +\frac{1}{2\xi_0}\partial_\mu A_\mu^{\prime a}(x)\partial_\mu A_\mu^{\prime a}(x)
   \right]
   +i\int d^Dx\,\Bar{\psi}'(x)\left(\Slash{\partial}-m_0\right)\psi'(x),
\label{eq:(3.23)}
\end{align}
where $g_0$, $m_0$, and~$\xi_0$ are bare parameters. For the gauge invariant
Wilson action~\eqref{eq:(3.1)}, we must assume a gauge invariant bare
action~$S$ without the gauge fixing term, i.e., $\xi_0\to\infty$; here, for a
later convenience, we have included a gauge fixing term. Also, for a later
illustration, we generalize the diffusion equations Eqs.~\eqref{eq:(3.3)}
and~\eqref{eq:(3.5)} to the
following:~\cite{Luscher:2010iy,Luscher:2011bx,Luscher:2013cpa}
\begin{align}
   \partial_tA_\mu'(t,x)&=D_\nu'F_{\nu\mu}'(t,x)
   +\alpha_0D_\mu'\partial_\nu A_\nu'(t,x),&
   A_\mu'(t=0,x)&=A_\mu'(x),
\label{eq:(3.24)}\\
   \partial_t\psi'(t,x)
   &=\left[D_\mu'D_\mu'-\alpha_0\partial_\mu A_\mu'(t,x)\right]\psi'(t,x),&
   \psi'(t=0,x)&=\psi'(x),
\notag\\
   \partial_t\Bar{\psi}'(t,x)
   &=\Bar{\psi}'(t,x)
   \left[\overleftarrow{D}_\mu'\overleftarrow{D}_\mu'
   +\alpha_0\partial_\mu A_\mu'(t,x)\right],&
   \Bar{\psi}'(t=0,x)&=\Bar{\psi}'(x),
\label{eq:(3.25)}
\end{align}
where the extra terms proportional to~$\alpha_0>0$ are introduced to make the
perturbative solution to the diffusion equation well-behaved.\footnote{%
These terms are known as the Zwanziger term in the context of stochastic
quantization~\cite{Zwanziger:1981kg,Baulieu:1981ec,Nakagoshi:1983rn}.}

Then, to evaluate the integral~\eqref{eq:(3.1)} in powers of fields $A_\mu^a$,
$\psi$, $\bar{\psi}$, one uses propagators defined by quadratic terms in the
exponent,\footnote{%
If we approximate the integral~\eqref{eq:(3.1)} to the quadratic order and
carry out the Gaussian integration, we obtain the Wilson action
\begin{align}
   S_\Lambda
   &=-\frac{1}{2}\int_pA_\mu^a(-p)A_\nu^b(p)
   \left[
   (p^2\delta_{\mu\nu}-p_\mu p_\nu)\frac{1}{g_0^2e^{-2tp^2}+p^2/\Lambda^{D-2}}
   +p_\mu p_\nu\frac{1}{\xi_0g_0^2e^{-2\alpha_0tp^2}+p^2/\Lambda^{D-2}}
   \right]
\notag\\
   &\qquad{}
   -\int_p\Bar{\psi}(-p)
   \frac{\Slash{p}+im_0}{Z_\psi(\Lambda)^2e^{-2tp^2}+i(\Slash{p}+im_0)/\Lambda}.
\label{eq:(3.26)}
\end{align}
This reproduces the Wilson action at tree level obtained perturbatively in a
gauge-fixed version of GFERG~\cite{Miyakawa:2021hcx,Miyakawa:2021wus}. In this
way, the integral representation~\eqref{eq:(3.1)} may be directly employed to
obtain the perturbative solution to the gauge-fixed version of
GFERG~\cite{Sonoda:2025yyy}.}
\begin{equation}
   \exp\left[
   -\frac{1}{2}\int_p A_\mu^{\prime a}(-p)K_{\mu\nu}^{ab}(p)A_\nu^{\prime b}(p)\right]
   \exp\left[-\int_p\Bar{\psi}'(-p)K(p)\psi'(p)\right],
\label{eq:(3.27)}
\end{equation}
where
\begin{align}
   &(K^{-1})_{\mu\nu}^{ab}(p,q)
\notag\\
   &=g_0^2\delta^{ab}
   \left[\left(\delta_{\mu\nu}-\frac{p_\mu p_\nu}{p^2}\right)
   \frac{1}{p^2+g_0^2\Lambda^{D-2}e^{-2tp^2}}
   +\frac{p_\mu p_\nu}{p^2}\frac{\xi_0}
   {p^2+\xi_0g_0^2\Lambda^{D-2}e^{-2\alpha_0tp^2}}\right]
   \delta(p+q),
\label{eq:(3.28)}
\end{align}
and
\begin{equation}
   (K^{-1})(p,q)=\frac{1}{\Slash{p}+im_0-i\Lambda Z_\psi(\Lambda)^2e^{-2tp^2}}
   \delta(p+q).
\label{eq:(3.29)}
\end{equation}
For the gauge invariant bare action~$S$ in~Eq.~\eqref{eq:(3.1)}, we take
$\alpha_0=0$ and~$\xi_0=+\infty$, and the propagator~\eqref{eq:(3.28)} becomes
\begin{align}
   &(K^{-1})_{\mu\nu}^{ab}(p,q)
\notag\\
   &=g_0^2\delta^{ab}
   \left[\left(\delta_{\mu\nu}-\frac{p_\mu p_\nu}{p^2}\right)
   \frac{1}{p^2+g_0^2\Lambda^{D-2}e^{-2tp^2}}
   +\frac{p_\mu p_\nu}{p^2}\frac{1}
   {g_0^2\Lambda^{D-2}}\right]
   \delta(p+q).
\label{eq:(3.30)}
\end{align}
This propagator, analogous to the massive vector field (the Proca field),
exhibits a very bad UV behavior; the longitudinal part approaches a constant
as~$|p|\to\infty$. This behavior worsens the convergence property of momentum
integrals appearing in the perturbative expansion of the Wilson
action~$S_\Lambda$. We have indeed encountered a diverging momentum integral
corresponding to this behavior in a one-loop calculation of the fermion kinetic
term in the gauge invariant Wilson action of QED~\cite{Sonoda:2025xxx}. In this
way, it is quite conceivable that the perturbative solution to the gauge
invariant GFERG~\eqref{eq:(3.14)} is ill-defined.\footnote{%
Even with the diffusion equations modified as Eqs.~\eqref{eq:(3.24)}
and~\eqref{eq:(3.25)}, the Wilson action~\eqref{eq:(3.1)} is gauge invariant if
the bare action~$S$ is gauge invariant~\cite{Sonoda:2020vut}. In such a
formulation, the longitudinal part of~Eq.~\eqref{eq:(3.28)} exponentially
grows as~$\sim e^{2\alpha_0 tp^2}$. Gauge invariance on the other hand implies
corresponding $\alpha_0$ dependence of the interaction vertices. For example,
the interaction vertex with two longitudinal gauge fields gets multiplied
by~$e^{-2\alpha_0t p^2}$. The resulting UV divergences remain either power like or
logarithmic. We can confirm this divergent behavior in the perturbative
calculation of the fermion kinetic term in the Wilson action of
QED~\cite{Miyakawa:2021wus}.}

We believe, nevertheless, that the gauge invariant GFERG
equation~\eqref{eq:(3.14)} itself is meaningful nonperturbatively. This belief
comes from the connection to the gradient flow formalism which can be
nonperturbatively defined by lattice regularization without gauge fixing. In
possible nonperturbative applications, the remarkable property of GFERG
preserving manifest gauge invariance should be quite helpful. We hope to
consider a nonperturbative application of GFERG equation~\eqref{eq:(3.14)} in
the near future.

\section{Gauge fixed version of GFERG with the NL field}
\label{sec:4}
\subsection{Formulation}
\label{sec:4.1}
In the previous section, we have observed that the perturbative expansion of
the gauge invariant Wilson action exhibits a quite bad UV behavior. This
prompts us to formulate a gauge fixed version of GFERG by introducing a gauge
fixing term into the bare action~$S$; it must be rather BRST invariant
containing the FP ghost-anti-ghost fields and the NL field.

We define a Wilson action~$S_\Lambda$ containing the FP ghost field~$c$,
anti-ghost field~$\Bar{c}$, and the NL field~$B$ by\footnote{%
The Gaussian integrand for the $B$ field has (as is usual for an auxiliary
field) a wrong sign. We assume that a certain prescription is in place to take
care of this point. For instance, it could be possible to make sense of the
integral by analytic continuation.}
\begin{align}
   &e^{S_\Lambda[A,\Bar{c},c,B,\Bar{\psi},\psi]}
\notag\\
   &:=\mathcal{N}(\Lambda)
   \int[dA'][dc'][d\Bar{c}'][dB'][d\psi'][d\Bar{\psi}']\,
\notag\\
   &\qquad\qquad{}
   \times\exp\left\{
   -\frac{\Lambda^{D-2}}{2}\int d^Dx\,
   \left[
   A_\mu^a(x)-A_\mu^{\prime a}(t,x)
   \right]^2\right\}
\notag\\
   &\qquad\qquad{}
   \times\exp\left\{
   -\Lambda^{D-2}\int d^Dx\,
   \left[\Bar{c}^a(x)-Z_{\Bar{c}}(\Lambda)\Bar{c}^{\prime a}(t,x)\right]
   \left[c^a(x)-c^{\prime a}(t,x)\right]
   \right\}
\notag\\
   &\qquad\qquad{}
   \times\exp\left\{
   \frac{\Lambda^{D-4}}{2}\int d^Dx\,
   \left[
   B^a(x)-Z_{\Bar{c}}(\Lambda)B^{\prime a}(t,x)
   \right]^2\right\}
\notag\\
   &\qquad\qquad{}
   \times\exp\left\{
   i\Lambda\int d^Dx\,
   \left[\Bar{\psi}(x)-Z_\psi(\Lambda)\Bar{\psi}'(t,x)\right]
   \left[\psi(x)-Z_\psi(\Lambda)\psi'(t,x)\right]
   \right\}
\notag\\
   &\qquad\qquad\qquad{}
   \times e^{S[A',\Bar{c}',c',B',\Bar{\psi}',\psi']},
\label{eq:(4.1)}
\end{align}
where
\begin{align}
   \mathcal{N}(\Lambda)
   &:=\biggl\{
   \int[dA][dc][d\Bar{c}][dB][d\psi][d\Bar{\psi}]\,
\notag\\
   &\qquad\qquad{}
   \times
   \exp\left[
   -\frac{\Lambda^{D-2}}{2}\int d^Dx\,A_\mu^a(x)^2\right]
   \exp\left[
   -\Lambda^{D-2}\int d^Dx\,\Bar{c}^a(x)c^a(x)\right]
\notag\\
   &\qquad\qquad{}
   \times
   \exp\left[
   \frac{\Lambda^{D-4}}{2}\int d^Dx\,B^a(x)^2\right]
   \exp\left[
   i\Lambda\int d^Dx\,\Bar{\psi}(x)\psi(x)\right]
   \biggr\}^{-1}.
\label{eq:(4.2)}
\end{align}
In this definition, the diffusion equations for the gauge and fermion fields
are given by Eqs.~\eqref{eq:(3.24)} and~\eqref{eq:(3.25)}, respectively. The
diffusion equation for the ghost field is set to be\footnote{%
Our $c'(t,x)$ thus corresponds to the flowed ghost field~$d(t,x)$
in~Ref.~\cite{Luscher:2011bx}.}
\begin{equation}
   \partial_tc^{\prime a}(t,x)=\alpha_0D_\mu'\partial_\mu c^{\prime a}(t,x),\qquad
   c^{\prime a}(t=0,x)=c^{\prime a}(x).
\label{eq:(4.3)}
\end{equation}

In~Eq.~\eqref{eq:(4.1)}, we also diffuse the NL field~$B(x)$ and the anti-ghost
field~$\Bar{c}(x)$ by the free diffusion,
\begin{align}
   \partial_t B^{\prime a}(t,x)
   &=\beta_0\partial_\mu\partial_\mu B^{\prime a}(t,x),\qquad
   B^{\prime a}(t=0,x)=B^{\prime a}(x),
\notag\\
   \partial_t\Bar{c}^{\prime a}(t,x)
   &=\beta_0\partial_\mu\partial_\mu\Bar{c}^{\prime a}(t,x),\qquad
   \Bar{c}^{\prime a}(t=0,x)=\Bar{c}^{\prime a}(x),
\label{eq:(4.4)}
\end{align}
with a constant~$\beta_0>0$. The reason for this prescription is explained
later.

Then, by now a familiar procedure, the following equality of the generating
functionals follows from~Eq.~\eqref{eq:(4.1)}:
\begin{align}
   &\int[dA][dc][d\Bar{c}][dB][d\psi][d\Bar{\psi}]\,
   e^{S_\Lambda[A,\Bar{c},c,B,\Bar{\psi}',\psi']}
\notag\\
   &\qquad{}
   \times
   \exp\left[
   -\frac{1}{2\Lambda^{D-2}}\int d^Dx\,
   \frac{\delta^2}{\delta A_\mu^a(x)\delta A_\mu^a(x)}\right]
   \exp\left[
   \frac{1}{\Lambda^{D-2}}\int d^Dx\,
   \frac{\delta}{\delta c^a(x)}\frac{\delta}{\delta\Bar{c}^a(x)}\right]
\notag\\
   &\qquad{}
   \times
   \exp\left[
   \frac{1}{2\Lambda^{D-4}}\int d^Dx\,
   \frac{\delta^2}{\delta B^a(x)\delta B^a(x)}\right]
   \exp\left[
   \frac{i}{\Lambda}\int d^Dx\,
   \frac{\delta}{\delta\psi(x)}\frac{\delta}{\delta\Bar{\psi}(x)}\right]
\notag\\   
   &\qquad{}
   \times\exp\biggl\{
   \int d^Dx\,J_\mu^a(x)A_\mu^a(x)
   +\int d^Dx\,\left[
   \Bar{\eta}_c^a(x)c^a(x)
   +\Bar{c}^a(x)\eta_{\Bar{c}}^a(x)
   \right]
\notag\\
   &\qquad\qquad\qquad{}
   +\int d^Dx\,J_B^a(x)B^a(x)
   +\int d^Dx\,\left[
   \Bar{\eta}(x)\psi(x)+\Bar{\psi}(x)\eta(x)\right]
   \biggr\}
\notag\\
   &=
   \int[dA'][dc'][d\Bar{c}'][dB'][d\psi'][d\Bar{\psi}']\,
   e^{S[A',\Bar{c}',c',B',\Bar{\psi}',\psi']}
\notag\\
   &\qquad{}
   \times\exp\biggl\{
   \int d^Dx\,J_\mu^a(x)A_\mu^{\prime a}(t,x)
   +\int d^Dx\,\left[
   \Bar{\eta}_c^a(x)c^{\prime a}(t,x)
   +Z_{\Bar{c}}(\Lambda)\Bar{c}^{\prime a}(t,x)\eta_{\Bar{c}}^a(x)
   \right]
\notag\\
   &\qquad\qquad\qquad{}
   +\int d^Dx\,J_B^a(x)Z_{\Bar{c}}(\Lambda)B^{\prime a}(t,x)
\notag\\
   &\qquad\qquad\qquad{}
   +\int d^Dx\,\left[
   \Bar{\eta}(x)Z_\psi(\Lambda)\psi'(t,x)
   +Z_\psi(\Lambda)\Bar{\psi}'(t,x)\eta(x)\right]
   \biggr\}.
\label{eq:(4.5)}
\end{align}
This gives the equality of correlation functions
\begin{align}
   &\biggl\langle
   \exp\left[
   -\frac{1}{2\Lambda^{D-2}}\int d^Dx\,
   \frac{\delta^2}{\delta A_\mu^a(x)\delta A_\mu^a(x)}\right]
   \exp\left[
   \frac{1}{\Lambda^{D-2}}\int d^Dx\,
   \frac{\delta}{\delta c^a(x)}\frac{\delta}{\delta\Bar{c}^a(x)}\right]
\notag\\
   &\qquad{}
   \times
   \exp\left[
   \frac{1}{2\Lambda^{D-4}}\int d^Dx\,
   \frac{\delta^2}{\delta B^a(x)\delta B^a(x)}\right]
\notag\\
   &\qquad{}
   \times
   A_{\mu_1}^{a_1}(x_1)\dotsb A_{\mu_N}^{a_N}(x_N)
   c^{b_1}(y_1)\dotsb c^{b_M}(y_M)
   \Bar{c}^{c_1}(z_1)\dotsb\Bar{c}^{c_L}(z_L)
   B^{d_1}(w_1)\dotsb B^{d_P}(w_P)
   \biggr\rangle_{S_\Lambda}
\notag\\
   &=Z_{\Bar{c}}(\Lambda)^{L+P}\biggl\langle
   A_{\mu_1}^{a_1}(t,x_1)\dotsb A_{\mu_N}^{a_N}(t,x_N)
   c^{b_1}(t,y_1)\dotsb c^{b_M}(t,y_M)
\notag\\
   &\qquad\qquad\qquad\qquad{}
   \times
   \Bar{c}^{c_1}(t,z_1)\dotsb\Bar{c}^{c_L}(t,z_L)
   B^{d_1}(t,w_1)\dotsb B^{d_P}(t,w_P)
   \biggr\rangle_S.
\label{eq:(4.6)}
\end{align}
We have omitted fermion fields for simplicity. This relation is the idea
underlying the definition~\eqref{eq:(4.1)}: the modified correlation function
computed with the Wilson action~\eqref{eq:(4.1)} is identical to the
correlation function of diffused or flowed fields computed with the bare
action~$S$. $Z_{\Bar{c}}(\Lambda)$ on the right-hand side is the conventional
common wave function renormalization factor of the anti-ghost field~$\Bar{c}$
and the NL field~$B$. From the analysis of~Ref.~\cite{Luscher:2011bx}, we know
that the correlation function on the right-hand side of~Eq.~\eqref{eq:(4.6)} is
finite (to all orders in perturbation theory) as~$\Lambda_0\to\infty$ under the
conventional renormalization of the coupling constants. In particular, neither
the diffused or flowed gauge field~$A_\mu(t,x)$ nor the diffused or flowed ghost
field~$c(t,x)$ requires any wave function renormalization. Then, the
equality~\eqref{eq:(4.6)} ensures that the correlation function on the
left-hand side is finite (at least to all orders in perturbation theory) under
the coupling constant renormalization which makes $S_\Lambda$ finite in the
continuum limit~$\Lambda_0\to\infty$. This finiteness holds even for the
products of elementary fields with identical spacetime coordinates; for the
gauge field~$A_\mu(t,x)$, the ghost field~$c(t,x)$ and the anti-ghost
field~$\Bar{c}(t,x)$, this finiteness of field products follows from the
argument in~\cite{Luscher:2011bx}. For the NL field~$B(t,x)$, the diffusion
in~Eq.~\eqref{eq:(4.4)} ensures this finiteness.

It has been shown in~\cite{Luscher:2010iy} that the correlation functions of
gauge invariant operators of the flowed gauge field (which does not contain the
flow time derivative) is independent of the parameter~$\alpha_0>0$. Thus the
choice of~$\alpha_0$ should not affect physics.

\subsection{BRST symmetry}
\label{sec:4.2}
In Eq.~\eqref{eq:(4.1)}, the bare action~$S$ is no longer gauge invariant, and
Eq.~\eqref{eq:(3.11)} does not hold anymore. Instead, we may derive WT identity
for the Wilson action~$S_\Lambda$ assuming the BRST invariance of the bare
action~$S$. Thus, let us consider a BRST type change of integration variables
in~Eq.~\eqref{eq:(4.1)}:
\begin{align}
   \delta A_\mu^{\prime a}(x)&=\eta D_\mu'c^{\prime a}(x),&
   \delta c^{\prime a}(x)&=-\eta\frac{1}{2}f^{abc}c^{\prime b}(x)c^{\prime c}(x),
\notag\\
   \delta\Bar{c}^{\prime a}(x)&=\eta B^{\prime a}(x),&
   \delta B^{\prime a}(x)&=0,
\notag\\
   \delta\psi'(x)&=-\eta c'(x)\psi'(x),&
   \delta\Bar{\psi}'(x)&=\Bar{\psi}'(x)\eta c'(x),
\label{eq:(4.7)}
\end{align}
where we have introduced an anticommuting constant~$\eta$ to keep the
statistics of the fields. We assume that the bare action~$S$ be invariant under
these. A crucially important point here is, as one can readily confirm, that
the BRST transformation given by~\eqref{eq:(4.7)} and the diffusion given
by~Eqs.~\eqref{eq:(3.24)}, \eqref{eq:(3.25)}, and~\eqref{eq:(4.3)} commute with
each other; this is also the case for~Eq.~\eqref{eq:(4.4)}. This implies that
the variation of initial values of the diffusion in~Eq.~\eqref{eq:(4.7)}
induces the BRST variation on the solution to the diffusion equations:
\begin{align}
   \delta A_\mu^{\prime a}(t,x)&=\eta D_\mu'c^{\prime a}(t,x),&
   \delta c^{\prime a}(t,x)
   &=-\eta\frac{1}{2}f^{abc}c^{\prime b}(t,x)c^{\prime c}(t,x),
\notag\\
   \delta\Bar{c}^{\prime a}(t,x)&=\eta B^{\prime a}(t,x),&
   \delta B^{\prime a}(t,x)&=0,
\notag\\
   \delta\psi'(t,x)&=-\eta c'(t,x)\psi'(t,x),&
   \delta\Bar{\psi}'(t,x)&=\Bar{\psi}'(t,x)\eta c'(t,x).
\label{eq:(4.8)}
\end{align}
Using this, from~Eq.~\eqref{eq:(4.1)} we obtain
\begin{align}
   0&=\int[dA'][dc'][d\Bar{c}'][dB'][d\psi'][d\Bar{\psi}']\,
\notag\\
   &\qquad{}
   \times\exp\left\{
   -\frac{\Lambda^{D-2}}{2}\int d^Dx\,
   \left[
   A_\mu^a(x)-A_\mu^{\prime a}(t,x)
   \right]^2\right\}
\notag\\
   &\qquad{}
   \times\exp\left\{
   -\Lambda^{D-2}\int d^Dx\,
   \left[\Bar{c}^a(x)-Z_{\Bar{c}}(\Lambda)\Bar{c}^{\prime a}(t,x)\right]
   \left[c^a(x)-c^{\prime a}(t,x)\right]
   \right\}
\notag\\
   &\qquad{}
   \times\exp\left\{
   \frac{\Lambda^{D-4}}{2}\int d^Dx\,
   \left[
   B^a(x)-Z_{\Bar{c}}(\Lambda)B^{\prime a}(t,x)
   \right]^2\right\}
\notag\\
   &\qquad{}
   \times\exp\left\{
   i\Lambda\int d^Dx\,
   \left[\Bar{\psi}(x)-Z_\psi(\Lambda)\Bar{\psi}'(t,x)\right]
   \left[\psi(x)-Z_\psi(\Lambda)\psi'(t,x)\right]
   \right\}
\notag\\
   &\qquad{}
   \times
   e^{S[A',\Bar{c}',c',B',\Bar{\psi}',\psi']}
\notag\\
   &\qquad\qquad{}
   \times
   \int d^Dx\,\biggl\{
   \Lambda^{D-2}\left[
   A_\mu^a(x)-A_\mu^{\prime a}(t,x)
   \right]\eta D_\mu'c^{\prime a}(t,x)
\notag\\
   &\qquad\qquad\qquad\qquad\qquad{}
   +\Lambda^{D-2}\left[\Bar{c}^a(x)-Z_{\Bar{c}}(\Lambda)\Bar{c}^{\prime a}(t,x)\right]
   (-\eta)\frac{1}{2}f^{abc}c^{\prime b}(t,x)c^{\prime c}(t,x)
\notag\\
   &\qquad\qquad\qquad\qquad\qquad{}
   +\Lambda^{D-2}\eta Z_{\Bar{c}}(\Lambda)B^{\prime a}(t,x)
   \left[c^a(x)-c^{\prime a}(t,x)\right]
\notag\\
   &\qquad\qquad\qquad\qquad\qquad{}
   -i\Lambda 
   Z_\psi(\Lambda)\Bar{\psi}'(t,x)\eta c'(t,x)
   \left[\psi(x)-Z_\psi(\Lambda)\psi'(t,x)\right]
\notag\\
   &\qquad\qquad\qquad\qquad\qquad{}
   -i\Lambda
   \left[\Bar{\psi}(x)-Z_\psi(\Lambda)\Bar{\psi}'(t,x)\right]
   Z_\psi(\Lambda)(-\eta)c'(t,x)\psi'(t,x)
   \biggr\}.
\label{eq:(4.9)}
\end{align}
This can be succinctly written as the WT identity
\begin{align}
   &\int d^Dx\,
   \biggl\{
   \frac{\delta}{\delta A_\mu^a(x)}\Hat{D}_\mu\Hat{c}^a(x)
   -\frac{\delta}{\delta c^a(x)}\frac{1}{2}f^{abc}\Hat{c}^b(x)\Hat{c}^c(x)
   +\frac{\delta}{\delta\Bar{c}^a(x)}\Hat{B}^a(x)
\notag\\
   &\qquad\qquad{}
   -\frac{\delta}{\delta\psi(x)}\Hat{c}(x)\Hat{\psi}(x)
   -\tr\left[
   \frac{\delta}{\delta\Bar{\psi}(x)}\Hat{\Bar{\psi}}(x)\Hat{c}(x)\right]
   \biggr\}\,e^{S_\Lambda[A,\Bar{c},c,B,\Bar{\psi},\psi]}=0,
\label{eq:(4.10)}
\end{align}
where $\Hat{c}^a(x)$ and~$\Hat{B}^a(x)$ denote the combinations
\begin{equation}
   \Hat{c}^a(x):=c^a(x)+\frac{1}{\Lambda^{D-2}}
   \frac{\delta}{\delta\Bar{c}^a(x)},
   \qquad
   \Hat{B}^a(x):=
   B^a(x)-\frac{1}{\Lambda^{D-4}}\frac{\delta}{\delta B^a(x)},
\label{eq:(4.11)}
\end{equation}
and $\Hat{A}_\mu^a(x)$ in~$\Hat{D}$, $\Hat{\psi}(x)$, and~$\Hat{\Bar{\psi}}(x)$
have been defined in~Eq.~\eqref{eq:(3.15)}. This WT identity is a consequence
of the BRST invariance of the bare action~$S$. Unfortunately, for non-Abelian
gauge theory, this equation is cubic in the Wilson action~$S_\Lambda$, and it is
inconceivable to find a nontrivial combination that satisfies this WT identity
exactly. Nevertheless, perturbation theory will work.

\subsection{GFERG equation}
\label{sec:4.3}
From Eq.~\eqref{eq:(4.1)}, we derive the GFERG equation
\begin{align}
   &-\Lambda\frac{\partial}{\partial\Lambda}
   e^{S_\Lambda[A,\Bar{c},c,B,\Bar{\psi},\psi]}
\notag\\
   &=\int d^Dx\,
   \biggl(
   \frac{\delta}{\delta A_\mu^a(x)}
   \left\{
   -\frac{2}{\Lambda^2}
   \left[
   \Hat{D}_\nu \Hat{F}_{\nu\mu}^a(x)
   +\alpha_0\Hat{D}_\mu\partial_\nu\Hat{A}_\nu^a(x)
   \right]
   +[(D-2)/2]\frac{1}{\Lambda^{D-2}}\frac{\delta}{\delta A_\mu^a(x)}\right\}
\notag\\
   &\qquad\qquad\qquad{}
   +\frac{\delta}{\delta c^a(x)}
   \left[
   \frac{2}{\Lambda^2}
   \alpha_0\Hat{D}_\mu\partial_\mu\Hat{c}^a(x)
   -\frac{D-2}{\Lambda^{D-2}}
   \frac{\delta}{\delta\Bar{c}^a(x)}\right]
\notag\\
   &\qquad\qquad\qquad{}
   +\frac{\delta}{\delta\Bar{c}^a(x)}
   \left[\frac{2}{\Lambda^2}\beta_0\partial_\mu\partial_\mu
   +\gamma_{\Bar{c}}(\Lambda)\right]\Hat{\Bar{c}}^a(x)
\notag\\
   &\qquad\qquad\qquad{}
   -\frac{\delta}{\delta B^a(x)}
   \left[\frac{2}{\Lambda^2}\beta_0\partial_\mu\partial_\mu
   +\gamma_{\Bar{c}}(\Lambda)\right]\Hat{B}^a(x)
\notag\\
   &\qquad\qquad\qquad{}
   +\frac{\delta}{\delta\Bar{\psi}(x)}\cdot
   \Hat{\Bar{\psi}}(x)
   \left\{
   \frac{2}{\Lambda^2}
   \left[\Hat{\overleftarrow{D}}_\mu\Hat{\overleftarrow{D}}_\mu
   +\alpha_0\partial_\mu\Hat{A}_\mu(x)\right]
   +\gamma_\psi(\Lambda)\right\}
\notag\\
   &\qquad\qquad\qquad{}
   +\frac{\delta}{\delta\psi(x)}
   \left\{
   \frac{2}{\Lambda^2}
   \left[\Hat{D}_\mu\Hat{D}_\mu-\alpha_0\partial_\mu\Hat{A}_\mu(x)\right]
   +\gamma_\psi(\Lambda)\right\}
   \Hat{\psi}(x)
\notag\\
   &\qquad\qquad\qquad{}
   -\frac{i}{\Lambda}\frac{\delta}{\delta\psi(x)}
   \frac{\delta}{\delta\Bar{\psi}(x)}
   \biggr)
   e^{S_\Lambda[A,\Bar{c},c,B,\Bar{\psi},\psi]},
\label{eq:(4.12)}
\end{align}
where we have introduced the anomalous dimension
\begin{equation}
   \gamma_{\Bar{c}}(\Lambda):=-\Lambda\frac{d\ln Z_{\Bar{c}}(\Lambda)}{d\Lambda}
\label{eq:(4.13)}
\end{equation}
and
\begin{equation}
   \Hat{\Bar{c}}^a(x):=
   \Bar{c}^a(x)-\frac{1}{\Lambda^{D-2}}\frac{\delta}{\delta c^a(x)}.
\label{eq:(4.14)}
\end{equation}
By construction, the evolution by this equation preserves the WT
identity~\eqref{eq:(4.10)}.

This completes the construction of a gauge-fixed version of GFERG with the NL
field~$B^a(x)$. This formulation allows perturbative analyses of the Wilson
action. A preliminary perturbative analysis to lower orders indicates that the
dependence of the Wilson action on the NL field~$B^a(x)$ can be quite
nontrivial even if the bare action is quadratic on~$B^a(x)$. In the next
section, we thus develop another gauge-fixed version of GFERG that does not
contain the NL field as another option.

\section{Gauge-fixed version of GFERG without the NL field}
\label{sec:5}
As a Wilson action~$S_\Lambda$ which contains the FP ghost field~$c$ and the
anti-ghost field~$\Bar{c}$, but not the NL field, we set
\begin{align}
   &e^{S_\Lambda[A,\Bar{c},c,\Bar{\psi},\psi]}
\notag\\
   &:=\mathcal{N}(\Lambda)
   \int[dA'][dc'][d\Bar{c}'][d\psi'][d\Bar{\psi}']\,
\notag\\
   &\qquad\qquad\qquad{}
   \times\exp\left\{
   -\frac{\Lambda^{D-2}}{2}\int d^Dx\,
   \left[
   A_\mu^a(x)-A_\mu^{\prime a}(t,x)
   \right]^2\right\}
\notag\\
   &\qquad\qquad\qquad{}
   \times\exp\left\{
   -\Lambda^{D-2}\int d^Dx\,
   \left[\Bar{c}^a(x)-Z_{\Bar{c}}(\Lambda)\Bar{c}^{\prime a}(x)\right]
   \left[c^a(x)-c^{\prime a}(t,x)\right]
   \right\}
\notag\\
   &\qquad\qquad\qquad{}
   \times\exp\left\{
   i\Lambda\int d^Dx\,
   \left[\Bar{\psi}(x)-Z_\psi(\Lambda)\Bar{\psi}'(t,x)\right]
   \left[\psi(x)-Z_\psi(\Lambda)\psi'(t,x)\right]
   \right\}
\notag\\
   &\qquad\qquad\qquad{}
   \times e^{S[A',\Bar{c}',c',\Bar{\psi}',\psi']},
\label{eq:(5.1)}
\end{align}
where
\begin{align}
   \mathcal{N}(\Lambda)
   &:=\biggl\{
   \int[dA][dc][d\Bar{c}][d\psi][d\Bar{\psi}]\,
\notag\\
   &\qquad\qquad{}
   \times
   \exp\left[
   -\frac{\Lambda^{D-2}}{2}\int d^Dx\,A_\mu^a(x)^2\right]
   \exp\left[
   -\Lambda^{D-2}\int d^Dx\,\Bar{c}^a(x)c^a(x)\right]
\notag\\
   &\qquad\qquad{}
   \times
   \exp\left[
   i\Lambda\int d^Dx\,\Bar{\psi}(x)\psi(x)\right]
   \biggr\}^{-1}.
\label{eq:(5.2)}
\end{align}
We obtain the Wilson action without the auxiliary field by integrating out the
auxiliary field: $e^{S_\Lambda[A,\bar{c},c,\bar{\psi},\psi]}=%
\int[dB]\,e^{S_ \Lambda[A,\bar{c},c,B,\bar{\psi},\psi]}$. The diffusion equations for
fields are given by~Eqs.~\eqref{eq:(3.24)}, \eqref{eq:(3.25)},
and~\eqref{eq:(4.3)}, respectively. From~Eq.~\eqref{eq:(5.1)}, we have the
equality of correlation functions
\begin{align}
   &\biggl\langle
   \exp\left[
   -\frac{1}{2\Lambda^{D-2}}\int d^Dx\,
   \frac{\delta^2}{\delta A_\mu^a(x)\delta A_\mu^a(x)}\right]
   \exp\left[
   \frac{1}{\Lambda^{D-2}}\int d^Dx\,
   \frac{\delta}{\delta c^a(x)}\frac{\delta}{\delta\Bar{c}^a(x)}\right]
\notag\\
   &\qquad{}
   \times
   A_{\mu_1}^{a_1}(x_1)\dotsb A_{\mu_N}^{a_N}(x_N)
   c^{b_1}(y_1)\dotsb c^{b_M}(y_M)
   \Bar{c}^{c_1}(z_1)\dotsb\Bar{c}^{c_L}(z_L)
   \biggr\rangle_{S_\Lambda}
\notag\\
   &=Z_{\Bar{c}}(\Lambda)^L\left\langle
   A_{\mu_1}^{a_1}(t,x_1)\dotsb A_{\mu_N}^{a_N}(t,x_N)
   c^{b_1}(t,y_1)\dotsb c^{b_M}(t,y_M)
   \Bar{c}^{c_1}(z_1)\dotsb\Bar{c}^{c_L}(z_L)
   \right\rangle_S.
\label{eq:(5.3)}
\end{align}
Here, we have omitted fermion fields again for simplicity. Almost the same
remarks we have made on~Eq.~\eqref{eq:(4.6)}, especially the finiteness of both
sides, can be repeated.

In Eq.~\eqref{eq:(5.1)}, assuming the BRST invariance of the bare action~$S$,
we may derive an WT identity for the Wilson action~$S_\Lambda$. We assume that
the bare action~$S$ is invariant under the following BRST type transformation:
\begin{align}
   \delta A_\mu^{\prime a}(x)&=\eta D_\mu'c^{\prime a}(x),&
   \delta c^{\prime a}(x)&=-\eta\frac{1}{2}f^{abc}c^{\prime b}(x)c^{\prime c}(x),
\notag\\
   \delta\Bar{c}^{\prime a}(x)
   &=\eta\frac{1}{\xi_0}\partial_\mu A_\mu^{\prime a}(x),&& 
\notag\\
   \delta\psi'(x)&=-\eta c'(x)\psi'(x),&
   \delta\Bar{\psi}'(x)&=\Bar{\psi}'(x)\eta c'(x),
\label{eq:(5.4)}
\end{align}
where $\xi_0$ stands for the bare gauge fixing parameter. As
in~Eq.~\eqref{eq:(4.7)}, since these transformations and the diffusion commute,
Eq.~\eqref{eq:(5.4)} induces the BRST variation on the solution to the
diffusion equations:
\begin{align}
   \delta A_\mu^{\prime a}(t,x)&=\eta D_\mu'c^{\prime a}(t,x),&
   \delta c^{\prime a}(t,x)
   &=-\eta\frac{1}{2}f^{abc}c^{\prime b}(t,x)c^{\prime c}(t,x),
\notag\\
   \delta\psi'(t,x)&=-\eta c'(t,x)\psi'(t,x),&
   \delta\Bar{\psi}'(t,x)&=\Bar{\psi}'(t,x)\eta c'(t,x).
\label{eq:(5.5)}
\end{align}
Using this fact, from~Eq.~\eqref{eq:(5.1)}, we have
\begin{align}
   0&=\int[dA'][dc'][d\Bar{c}'][d\psi'][d\Bar{\psi}']\,
\notag\\
   &\qquad{}
   \times\exp\left\{
   -\frac{\Lambda^{D-2}}{2}\int d^Dx\,
   \left[A_\mu^a(x)-A_\mu^{\prime a}(t,x)\right]^2
   \right\}
\notag\\
   &\qquad{}
   \times\exp\left\{
   -\Lambda^{D-2}\int d^Dx\,
   \left[\Bar{c}^a(x)-Z_{\Bar{c}}(\Lambda)\Bar{c}^{\prime a}(x)\right]
   \left[c^a(x)-c^{\prime a}(t,x)\right]
   \right\}
\notag\\
   &\qquad{}
   \times\exp\left\{
   i\Lambda\int d^Dx\,
   \left[\Bar{\psi}(x)-Z_\psi(\Lambda)\Bar{\psi}'(t,x)\right]
   \left[\psi(x)-Z_\psi(\Lambda)\psi'(t,x)\right]
   \right\}
\notag\\
   &\qquad{}
   \times
   e^{S[A',\Bar{c}',c',\Bar{\psi}',\psi']}
\notag\\
   &\qquad\qquad{}
   \times
   \int d^Dx\,\biggl\{
   \Lambda^{D-2}\left[
   A_\mu^a(x)-A_\mu^{\prime a}(t,x)
   \right]\eta D_\mu'c^{\prime a}(t,x)
\notag\\
   &\qquad\qquad\qquad\qquad\qquad{}
   +\Lambda^{D-2}\left[\Bar{c}^a(x)-Z_{\Bar{c}}(\Lambda)\Bar{c}^{\prime a}(x)\right] 
   (-\eta)\frac{1}{2}f^{abc}c^{\prime b}(t,x)c^{\prime c}(t,x)x
\notag\\
   &\qquad\qquad\qquad\qquad\qquad{}
   +\Lambda^{D-2}
   \eta\frac{Z_{\Bar{c}}(\Lambda)}{\xi_0}\partial_\mu A_\mu^{\prime a}(x)
   \left[c^a(x)-c^{\prime a}(t,x)\right]
\notag\\
   &\qquad\qquad\qquad\qquad\qquad{}
   +i\Lambda
   Z_\psi(\Lambda)\Bar{\psi}'(t,x)(-\eta)c'(t,x)
   \left[\psi(x)-Z_\psi(\Lambda)\psi'(t,x)\right]
 \notag\\
   &\qquad\qquad\qquad\qquad\qquad{}
   +i\Lambda
   \left[\Bar{\psi}(x)-Z_\psi(\Lambda)\Bar{\psi}'(t,x)\right]
   Z_\psi(\Lambda)\eta c'(t,x)\psi'(t,x)
   \biggr\}.
\label{eq:(5.6)}
\end{align}
This can be written as
\begin{align}
   &\int d^Dx\,
   \biggl\{
   \frac{\delta}{\delta A_\mu^a(x)}\Hat{D}_\mu\Hat{c}^a(x)
   -\frac{\delta}{\delta c^a(x)}\frac{1}{2}f^{abc}\Hat{c}^b(x)\Hat{c}^c(x)
\notag\\
   &\qquad\qquad{}
   +\frac{\delta}{\delta\Bar{c}^a(x)}
   \frac{Z_{\Bar{c}}(\Lambda)}{\xi_0}\partial_\mu
   \mathcal{I}_\mu^a\left[\Hat{A};(t,x)\right]
\notag\\
   &\qquad\qquad{}
   -\frac{\delta}{\delta\psi(x)}\Hat{c}(x)\Hat{\psi}(x)
   -\tr\left[
   \frac{\delta}{\delta\Bar{\psi}(x)}\Hat{\Bar{\psi}}(x)\Hat{c}(x)\right]
   \biggr\}e^{S_\Lambda[A,\Bar{c},c,\Bar{\psi},\psi]}=0.
\label{eq:(5.7)}
\end{align}
The hatted quantities are defined in~Eqs.~\eqref{eq:(3.15)}
and~\eqref{eq:(4.11)}. To express $A_\mu^{\prime a}(x)$ in~Eq.~\eqref{eq:(5.6)},
we have introduced the inverse mapping of diffusion:
\begin{equation}
   \text{$A_\mu^{\prime a}(t,x)$ is the solution to~Eq.~\eqref{eq:(3.24)}}
   \Leftrightarrow
   A_\mu^{\prime a}(x)=\mathcal{I}_\mu^a\left[A'(t,x);(t,x)\right].
\label{eq:(5.8)}
\end{equation}
The combination
$[Z_{\Bar{c}}(\Lambda)/\xi_0]\partial_\mu\mathcal{I}_\mu^a[\Hat{A};(t,x)]$ should
be a renormalized combination because the other quantities
in~Eq.~\eqref{eq:(5.7)} are finite as~$\Lambda_0\to\infty$.

The WT identity~\eqref{eq:(5.7)} is a consequence of the BRST invariance of the
bare action~$S$. Unfortunately, for non-Abelian gauge theory, this equation is
infinite order in~$\Hat{A}$, and hence in the Wilson action~$S_\Lambda$; it is
inconceivable to find a nontrivial combination that satisfies this WT identity
exactly. Nevertheless perturbation theory will work.

The GFERG corresponding to~Eq.~\eqref{eq:(5.1)} is given by
\begin{align}
   &-\Lambda\frac{\partial}{\partial\Lambda}
   e^{S_\Lambda[A,\Bar{c},c,\Bar{\psi},\psi]}
\notag\\
   &=\int d^Dx\,
   \biggl(
   \frac{\delta}{\delta A_\mu^a(x)}
   \left\{
   -\frac{2}{\Lambda^2}
   \left[
   \Hat{D}_\nu \Hat{F}_{\nu\mu}^a(x)
   +\alpha_0\Hat{D}_\mu\partial_\nu\Hat{A}_\nu^a(x)
   \right]
   +[(D-2)/2]\frac{1}{\Lambda^{D-2}}\frac{\delta}{\delta A_\mu^a(x)}\right\}
\notag\\
   &\qquad\qquad\qquad{}
   +\frac{\delta}{\delta c^a(x)}
   \left[
   \frac{2}{\Lambda^2}
   \alpha_0\Hat{D}_\mu\partial_\mu\Hat{c}^a(x)
   -\frac{D-2}{\Lambda^{D-2}}
   \frac{\delta}{\delta\Bar{c}^a(x)}\right]
\notag\\
   &\qquad\qquad\qquad{}
   +\frac{\delta}{\delta\Bar{c}^a(x)}\gamma_{\Bar{c}}(\Lambda)\Hat{\Bar{c}}^a(x)
\notag\\
   &\qquad\qquad\qquad{}
   +\frac{\delta}{\delta\Bar{\psi}(x)}\cdot
   \Hat{\Bar{\psi}}(x)
   \left\{
   \frac{2}{\Lambda^2}
   \left[\Hat{\overleftarrow{D}}_\mu\Hat{\overleftarrow{D}}_\mu
   +\alpha_0\partial_\mu\Hat{A}_\mu(x)\right]
   +\gamma_\psi(\Lambda)\right\}
\notag\\
   &\qquad\qquad\qquad{}
   +\frac{\delta}{\delta\psi(x)}
   \left\{
   \frac{2}{\Lambda^2}
   \left[\Hat{D}_\mu\Hat{D}_\mu-\alpha_0\partial_\mu\Hat{A}_\mu(x)\right]
   +\gamma_\psi(\Lambda)\right\}
   \Hat{\psi}(x)
\notag\\
   &\qquad\qquad\qquad{}
   -\frac{i}{\Lambda}\frac{\delta}{\delta\psi(x)}
   \frac{\delta}{\delta\Bar{\psi}(x)}
   \biggr)
   e^{S_\Lambda[A,\Bar{c},c,\Bar{\psi},\psi]}.
\label{eq:(5.9)}
\end{align}
Note that the GFERG equation~\eqref{eq:(5.9)} itself does not depend on the
gauge fixing parameter~$\xi_0$; the dependence of~$S_\Lambda$ on~$\xi_0$ is
introduced by the WT relation~\eqref{eq:(5.7)}.

We may apply the present gauge-fixed GFERG formalism to QED. Noting that the
inverse map~\eqref{eq:(5.8)} in the Abelian gauge theory is simply given
by~$\mathcal{I}_\mu[A;(t,p)]=e^{+tp^2}A_\mu(p)$ in momentum space (here we have
set~$\alpha_0=1$), we find that the present gauge-fixed version of GFERG
coincides completely with the GFERG formulation studied
in~Ref.~\cite{Miyakawa:2021wus}. Thus, the formulation in this section provides
a natural generalization of the GFERG in~Ref.~\cite{Miyakawa:2021wus} to
non-Abelian theories. In~Ref.~\cite{Miyakawa:2021wus}, by perturbation theory,
we have found the same fermion anomalous dimension as in the gradient flow
formalism~\cite{Luscher:2013cpa}. This observation now has an explanation
because of the equality of the correlation functions in~Eq.~\eqref{eq:(5.3)}.

\section{1PI action}
\label{sec:6}
In the applications of ERG it is the 1PI Wilson
action~\cite{Morris:1993qb,Nicoll:1977hi,Wetterich:1992yh,Wetterich:1993ne,Bervillier:2008an}, as opposed to the Wilson action~$S_\Lambda$, that is often
studied. Thus, it should be useful to recapitulate how to define the 1PI action
in GFERG~\cite{Sonoda:2022fmk}.

Let $S_\Lambda [A,\bar{c},c,\bar{\psi},\psi]$ be the Wilson action discussed
in~Section~\ref{sec:5}. As noted in~Ref.~\cite{Sonoda:2022fmk}, the 1PI
action~$\mathit{\Gamma}_\Lambda$ can be obtained from the Wilson
action~$S_\Lambda$ by the Legendre transformation:
\begin{align}
   \mathit{\Gamma}_\Lambda[\mathcal{A},\Bar{C},C,\Bar{\Psi},\Psi]
   &:=\frac{\Lambda^{D-2}}{2}\int d^Dx\,
   \left[\mathcal{A}_\mu^a(x)-A_\mu^a(x)\right]^2
\notag\\
   &\qquad{}
   +\Lambda^{D-2}\int d^Dx\,
   \left[\Bar{C}^a(x)-\Bar{c}^a(x)\right]\left[C^a(x)-c^a(x)\right]
\notag\\
   &\qquad{}
   -i\Lambda\int d^Dx\,
   \left[\Bar{\Psi}(x)-\Bar{\psi}(x)\right]\left[\Psi(x)-\psi(x)\right]
\notag\\
   &\qquad{}
   +S_\Lambda[A,\Bar{c},c,\Bar{\psi},\psi],
\label{eq:(6.1)}
\end{align}
where
\begin{align}
   \mathcal{A}_\mu^a(x)
   &:=A_\mu^a(x)
   +\frac{1}{\Lambda^{D-2}}\frac{\delta S_\Lambda}{\delta A_\mu^a(x)},&&
\notag\\
   C^a(x)
   &:=c^a(x)+\frac{1}{\Lambda^{D-2}}
   \frac{\delta}{\delta\Bar{c}^a(x)}S_\Lambda,&
   \Bar{C}^a(x)
   &:=\Bar{c}^a(x)-\frac{1}{\Lambda^{D-2}}
   \frac{\delta}{\delta c^a(x)}S_\Lambda,&
\notag\\
   \Psi(x)&:=\psi(x)
   +\frac{i}{\Lambda}\frac{\delta}{\delta\Bar{\psi}(x)}S_\Lambda,&
   \Bar{\Psi}(x)&:=\Bar{\psi}(x)
   -\frac{i}{\Lambda}\frac{\delta}{\delta\psi(x)}S_\Lambda.&
\label{eq:(6.2)}
\end{align}
The inverse transformation gives
\begin{align}
   A_\mu^a(x)
   &=\mathcal{A}_\mu^a(x)
   -\frac{1}{\Lambda^{D-2}}
   \frac{\delta\mathit{\Gamma}_\Lambda}{\delta\mathcal{A}_\mu^a(x)},&&
\notag\\
   c^a(x)
   &=C^a(x)-\frac{1}{\Lambda^{D-2}}
   \frac{\delta}{\delta\Bar{C}^a(x)}\mathit{\Gamma}_\Lambda,&
   \Bar{c}^a(x)
   &=\Bar{C}^a(x)+\frac{1}{\Lambda^{D-2}}
   \frac{\delta}{\delta C^a(x)}\mathit{\Gamma}_\Lambda,&
\notag\\
   \psi(x)&=\Psi(x)
   -\frac{i}{\Lambda}\frac{\delta}{\delta\Bar{\Psi}(x)}\mathit{\Gamma}_\Lambda,&
   \Bar{\psi}(x)&=\Bar{\Psi}(x)
   +\frac{i}{\Lambda}\frac{\delta}{\delta\Psi(x)}\mathit{\Gamma}_\Lambda.&
\label{eq:(6.3)}
\end{align}
Hence, we obtain
\begin{align}
   \frac{\delta S_\Lambda}{\delta A_\mu^a(x)}
   &=\frac{\delta\mathit{\Gamma}_\Lambda}{\delta\mathcal{A}_\mu^a(x)},&&
\notag\\
   \frac{\delta}{\delta c^a(x)}S_\Lambda
   &=\frac{\delta}{\delta C^a(x)}\mathit{\Gamma}_\Lambda,&
   \frac{\delta}{\delta\Bar{c}^a(x)}S_\Lambda
   &=\frac{\delta}{\delta\Bar{C}^a(x)}\mathit{\Gamma}_\Lambda,
\notag\\
   \frac{\delta}{\delta\psi(x)}S_\Lambda
   &=\frac{\delta}{\delta\Psi(x)}\mathit{\Gamma}_\Lambda,&
   \frac{\delta}{\delta\Bar{\psi}(x)}S_\Lambda
   &=\frac{\delta}{\delta\Bar{\Psi}(x)}\mathit{\Gamma}_\Lambda.
\label{eq:(6.4)}
\end{align}
Using these relations, it is possible to rewrite the GFERG
equation~\eqref{eq:(5.9)} for the Wilson action to that for the 1PI action.
Such an explicit expression can be found for QED in~Ref.~\cite{Sonoda:2022fmk}.

\section{Conclusion}
\label{sec:7}
In this paper, we formulate GFERG, a variant of Wilson ERG, that preserves
gauge invariance maximally, on the basis of the Reuter formula. With this
formulation, we have clarified some points left unresolved in our previous
studies. We hope and believe that the formulations presented in this paper can
provide a solid basis for both perturbative and nonperturbative applications
of GFERG.

The ERG approach has the advantage that a theory is formulated in a continuous
spacetime as opposed to a discrete lattice. This gives us a hope that the
formulation can be made compatible with spacetime symmetries including
supersymmetry. For some time supersymmetric gradient flows have been
studied~\cite{Kikuchi:2014rla,Kadoh:2018qwg,Kadoh:2019glu,Kadoh:2022are,Kadoh:2023gqa,Kadoh:2023mof}.  We think it possible to formulate GFERG that is not only
gauge invariant but also supersymmetric. It should be of some interest to
pursue along this line.

\section*{Acknowledgments}
We would like to thank Nobuyuki Ishibashi for encouragement. We thank the RIKEN
Center for Interdisciplinary Theoretical and Mathematical Sciences (iTHEMS) for
providing convenience for the collaboration. The work of H.~Suzuki was
partially supported by JSPS KAKENHI Grant Number JP23K03418.

%\appendix



%\bibliographystyle{ptephy}
%\bibliography{sample}
%
% once the .bbl file has been generated then place the text in your article.

%% \vspace{0.2cm}
%% \noindent
%% For references, note how to include DOI information from examples below. 

%This is added by T. Yoneya (editor-in-chief) on 2020/07/09.

\let\doi\relax

%without this code before the command "\begin{thebibliography}{}" , an error will be %flagged. When the bibliography is provided as separate .bib file, then this code %should be placed above the commands "\bibliographystyle{}" and "\bibliography{}" %inside the main TeX file. 

%% \begin{thebibliography}{9}

%% \bibitem{1}
%% J. P.~Blaizot, and E.~Iancu, Phys. Rep. {\bf 359}, 355 (2002).
%% \doi{https://doi.org/10.1016/S0370-1573(01)00061-8}

%% \bibitem{2}
%% M.~Gyulassy, and L.~McLerran, Nucl.\ Phys.\  A {\bf 750}, 30 (2005). \\ \doi{https://doi.org/10.1016/j.nuclphysa.2004.10.034}

%% \bibitem{3}
%% S.~Aoki et al. [JLQCD Collaboration], Phys. Rev. D 72, 054510 (2005). \\
%% \doi{https://doi.org/10.1103/PhysRevD.72.05451}

%% \bibitem{4}
%% S.~Alekhin, A.~Djouadi, and S.~Moch, Phys. Lett. B 716, 214 (2012) [arXiv:1207.0980 [hep-ph]]. \doi{https://doi.org/10.1016/j.physletb.2012.08.024}

%% \end{thebibliography}

%\bibliographystyle{ptephy}
%\bibliography{anomaly}

%\clearpage

\end{document}